\documentclass[aps,pra,twocolumn,superscriptaddress]{revtex4-2}

\usepackage{amsmath, physics}
\usepackage{amssymb,color,xcolor}
\usepackage{graphicx,float,subfigure}
\usepackage{bm} 
\usepackage[colorlinks=true,linkcolor=blue,citecolor=blue,urlcolor=blue]{hyperref}

\begin{document}

	\title{Robust Entanglement Dynamics in Driven Open Quantum Systems}

	\author{Aqsa Mushtaq}
	\affiliation{Center for High Energy Physics, University of the Punjab}
	
	\author{Chaimae Banouni}
	\affiliation{Laboratory of Theoretical Physics, Particles, Modeling and Energies, Faculty of Sciences,\\
		Mohamed First University, Oujda, Morocco}

	\author{Mahboob Ul Haq}
	\affiliation{Department of Physics, University of Malakand, Dir Lower, Khyber Pakhtunkhwa, Pakistan}
	\affiliation{Govt. Post Graduate College Timergara, Khyber Pakhtunkhwa, Pakistan}

	\author{S. M. Zangi\^{*}}
	\affiliation{Department of Physics, University of Sargodha,
		 Sargodha 40100, Pakistan}
	\date{\today}

	\begin{abstract}
		We investigate the dynamics of key quantum correlations—Negativity (NG), Quantum Discord (QD), and Quantum-Memory-Assisted Entropic Uncertainty (QM-EUR)—in a bipartite two-qubit system under the influence of external pulses and various decoherence channels, including amplitude damping ($\gamma_{\rm amp}$), pure dephasing ($\gamma_{\rm deph}$), and pulse-induced dephasing ($G$), while different regimes of inter-qubit coupling ($J_{zz}, J_{xx}$), qubit energy splitting ($\varepsilon$), and pulse parameters ($A_{\rm pulse}, \beta_{\rm pulse}$) are explored. Our results show that inter-qubit coupling and energy splitting $\epsilon$ significantly influence the dynamics, producing pronounced oscillations in the weak-coupling regime and protecting pre-existing entanglement in the strong-coupling regime. NG is the most sensitive, QD persists longer revealing nonclassical correlations independent of entanglement, and QM-EUR reflects residual quantum memory and entropic uncertainty, showing that quantum signatures survive even when NG and QD are weak. Pulse amplitude and width effectively control the generation and dissipation of correlations, while the intensity of pulse-induced dephasing modulates the balance between sustained oscillations and rapid decoherence. The initial state also plays a crucial role: a partially entangled initial state is more resilient to perturbations, preserving correlations over time, whereas a separable state exhibits cycles of entanglement creation and destruction. Thus, by adjusting system parameters, it is possible to control the stability and lifetime of correlations and coherence, providing a framework to optimize quantum systems for applications requiring both strong entanglement and long-lasting coherence, such as quantum computing and secure communication.
		
	\end{abstract}
	
		\maketitle
		
	\newpage
	\section{Introduction}
	\label{sec:intro}

	Quantum computing harnesses fundamental quantum-mechanical features—superposition, entanglement, and coherence—to achieve computational and communication tasks beyond the reach of classical systems~\cite{ref:nielsen_chuang}. Realistic quantum devices, however, are intrinsically open systems whose unavoidable coupling to the environment leads to decoherence, dissipation, and noise~\cite{breuer2002theory}. Understanding and mitigating these effects remains a central challenge in the quest for scalable, fault-tolerant quantum technologies~\cite{terhal2015quantum,viola1998dynamical,temme2017error,glaser2015training}.
	
	Among the simplest yet most informative models for exploring these issues are coupled two-qubit systems, where local energy splittings compete with nonlocal exchange interactions such as the Ising $(\hat{\sigma}_z\!\otimes\!\hat{\sigma}_z)$ and XX $(\hat{\sigma}_x\!\otimes\!\hat{\sigma}_x)$ terms. Their interplay governs the creation and preservation of quantum correlations, reflecting the balance between coherent coupling and environmental decoherence~\cite{amico2008entanglement,blais2021circuit}. Engineering these couplings in controllable architectures—ranging from superconducting circuits to trapped ions—has enabled experimental access to regimes where entanglement, state transfer, and decoherence coexist on comparable timescales.
	
	Time-dependent driving fields and precisely shaped control pulses provide powerful means to manipulate such dynamics. By embedding tailored pulse envelopes into the system Hamiltonian, one can overcome static limitations, selectively generate entanglement, and even induce effective Hamiltonians that mimic desired interactions~\cite{martinis2014fast,caneva2011chopped,bialczak2011fast,eckardt2017colloquium}. Among smooth pulse shapes, the hyperbolic secant (\textit{sech}) profile offers adiabatic amplitude variation, reducing spectral leakage and enabling coherent population control. When applied to coupled qubits, such drives can also introduce correlated dephasing channels—features that can both preserve and suppress quantum correlations depending on their relative strength.
	
	In parallel, diverse measures have been developed to characterize quantum correlations beyond simple entanglement. \textit{Negativity} provides an efficient quantifier of bipartite entanglement~\cite{vidal2002computable}, while \textit{quantum discord} captures nonclassical correlations that may persist even in separable states~\cite{ollivier2001quantum,modi2012classical}. More recently, the \textit{quantum-memory-assisted entropic uncertainty relation} (QM–EUR) has emerged as a complementary probe of residual quantumness, connecting information-theoretic uncertainty with bipartite correlations~\cite{berta2010uncertainty}. Together, these measures offer a multidimensional view of quantum correlation dynamics under open-system evolution.
	
	In this work, we investigate the time evolution of \textbf{Negativity (NG)}, \textbf{Quantum Discord (QD)}, and \textbf{QM–EUR} in a driven two-qubit system interacting with a dissipative environment. The model includes both XX and ZZ couplings, external driving with a hyperbolic secant pulse, and three decoherence channels—amplitude damping $(\gamma_{\mathrm{amp}})$, pure dephasing $(\gamma_{\mathrm{deph}})$, and pulse-induced correlated dephasing $(G)$. Using the Lindblad master equation, we simulate how varying coupling strengths $(J_{xx}, J_{zz})$, qubit energy splitting $(\varepsilon)$, and pulse parameters $(A_{\mathrm{pulse}}, \beta_{\mathrm{pulse}})$ influence the interplay between coherence and decoherence. The results demonstrate that by tuning the drive parameters and coupling regime, one can effectively control the generation, decay, and revival of quantum correlations, providing insight into designing noise-resilient quantum operations and long-lived entanglement in realistic architectures.

	\section{SYSTEM DESCRIPTION AND HAMILTONIAN}
	\label{sec:NO_st}
	Defining the Hamiltonian of a two-qubit system is the first step in studying the impact of driving and dissipation. In a two-qubit system, denoted as qubits 0 and 1 (k=0,1), the dynamics are governed by a total Hamiltonian:
	\begin{equation}
		H(t) = H_0 + H_D(t)
	\end{equation}
	This Hamiltonian is composed of two key components. The first is the static Hamiltonian $H_0$, which is time-independent and encapsulates the inherent interactions between the two qubits, defining their natural behavior. The second is the driving Hamiltonian $H_D(t)$, a time-dependent component that accounts for external influences or control fields applied to the system. Together, these components dictate the coherent evolution of the two-qubit system, shaping its quantum behavior over time.\\
	A static Hamiltonian does not explicitly involve the time variable \cite{deClercq2016}. It shows the intrinsic energy splitting and inter-qubit interaction. Here, $\epsilon_k$ represents the energy level splitting for qubit k,  $\Delta_k$ denotes the static tunneling amplitude or fixed transverse field strength applied to qubit k (where parameters $\Delta_0$ = $\Delta_1$ = 0 and $J_{yy}$ = 0). Given Hamiltonian reduces to:
	\begin{equation}
		H_0 = \frac{\epsilon_0}{2} \, \sigma_z^{(0)} + \frac{\epsilon_1}{2} \, \sigma_z^{(1)} + J_{zz} \, \sigma_z^{(0)} \sigma_z^{(1)} + J_{xx} \, \sigma_x^{(0)} \sigma_x^{(1)}
	\end{equation}
	
	Furthermore, quantum operations are based on Pauli operators commonly referred to as sigma operators $\sigma_x$, $\sigma_y$, $\sigma_z$ along with the identity operator I. These operators signify X (Bit-flip operator), Y (Bit-phase-flip operator), and Z (Phase-flip operator) \cite{Vaicaitis_Rimas_Pauli_operators}. $J_{zz}$ is the Ising-type coupling strength along the ZZ and $J_{xx}$ along XX between the two qubits. Matrices represent these:
	\[
	\sigma_x = X =
	\begin{pmatrix}
		0 & 1 \\
		1 & 0
	\end{pmatrix}, \quad
	\sigma_z = Z =
	\begin{pmatrix}
		1 & 0 \\
		0 & -1
	\end{pmatrix}
	\]
	
	In the two-qubit Hilbert space, the operator $\sigma_z^{(0)}$ denotes $\sigma_z \otimes I$, and $\sigma_z^{(1)}$ denotes $I \otimes \sigma_z$ and the interaction terms like $\sigma_z^{(0)} \sigma_z^{(1)}$ are defined as $\sigma_z^{(0)} \sigma_z^{(1)}$ = $\sigma_z \otimes \sigma_z$. \\
	
	The idea of employing an external time-dependent field to affect the coherent dynamics of a quantum system has attracted a lot of theoretical and experimental interest from a variety of physics and chemical disciplines \cite{majeed2020quantumzenoantizenoeffects}. One popular tool, an external driving field is used for manipulating qubits and now clearly shows it drastically influence the temporal evolution of quantum systems \cite{Kofman_2001, Gordon_2008, Gordon2006Universal}. An external coherent drive is applied to both qubits. The drive operator, $H_{\text{D}}^{\text{op}}$ of Pauli-X and Pauli-Y operators on each qubit:
	\begin{equation}
		H_{\text{D}}^{\text{op}} = \sigma_x^{(0)} + \sigma_y^{(0)} + \sigma_x^{(1)}+ \sigma_y^{(1)}
	\end{equation}
	
	A hyperbolic secant (sech) pulse profile is used to model the amplitude of the external driving field. It is applied to the qubits as a smooth, localized function of time and gives smooth rise and fall \cite{vitanov2017stimulated}. The inclusion of both $\sigma_x$ and $\sigma_y$ components in $H_D^{\mathrm{op}}$ 
	represents a coherent control field with equal in-phase and 
	quadrature components acting on each qubit. Physically, this corresponds to a 
	rotating drive in the $xy$-plane with fixed phase $\phi=\pi/4$, analogous to 
	microwave control in superconducting qubits where $I$–$Q$ modulation generates 
	phase-stable rotations~\cite{motzoi2009simple,krantz2019quantum,clarke2008superconducting}. 
	This rotating-field form helps maintain transverse coherence during pulse interaction, 
	reducing phase-localized decoherence and supporting the robust entanglement revival 
	observed in our results.
	In terms of mathematics, the pulse is defined as:
	\begin{equation}
		f(t) = \cfrac{A_\text{pulse}}{cosh(\beta_\text{pulse}(t-t_0)}
	\end{equation}
	where $A_\text{pulse}$ is the maximum amplitude of the pulse, $\beta_\text{pulse}$ controls the temporal width or spread of the pulse (larger 
	$\beta$ means a narrower pulse), $t_o$ is the center time of the pulse, typically $t_o$ = $T_\text{max}/2$.\\
	Lastly, the operator $H_{\text{D}}^{\text{op}}$ is scaled by the time-dependent amplitude f(t) to create the whole time-dependent driving Hamiltonian:
	\begin{equation}
		H_D(t) = f(t) \cdot H_{\text{D}}^{\text{op}}
	\end{equation}
	
	\section{System Dynamics: Lindblad Master Equation}
	\label{sec:system-dynamics}
	
	The dynamics of the open quantum system are described by the Lindblad master equation, which governs the time evolution of the system's density matrix $\rho(t)$. This equation accounts for both the coherent evolution under the Hamiltonian $H(t)$ and the incoherent dissipative processes that arise from the interaction with the environment. The general form of the equation is given by:
	\begin{equation}
		\frac{d\rho}{dt} = -i[H(t), \rho] + \sum_j \mathcal{L}_j[\rho]
	\end{equation}
	The second term on the right, $\sum_j \mathcal{L}_j[\rho]$, represents the contribution from dissipation. The Lindblad superoperator, $\mathcal{L}_j[\rho]$, is defined for each collapse operator $L_j$ as:
	\begin{equation}
		\mathcal{L}_j[\rho] = \gamma_j \left( L_j \rho L_j^\dagger - \tfrac{1}{2} \{ L_j^\dagger L_j , \rho \} \right)
	\end{equation}
	where $\gamma_j$ is the decay rate and $\{A,B\} = AB + BA$ is the anticommutator. The system is subject to two types of decay channels: standard Markovian decay and pulse-induced dephasing. These channels describe weak, constant environmental noise. The system experiences both amplitude damping (relaxation) and pure dephasing, each with a constant decay rate.

	Amplitude damping, or relaxation, describes energy loss from the qubit to the environment. This is represented by the following collapse operators for qubits 0 and 1:
	\begin{equation}
		L^{(0)}_a = \sigma^{(0)}_- \equiv |0\rangle\langle 1|^{(0)}, \quad \text{with rate } \gamma_{\text{amp}}
	\end{equation}
	\begin{equation}
		L^{(1)}_a = \sigma^{(1)}_- \equiv |0\rangle\langle 1|^{(1)}, \quad \text{with rate } \gamma_{\text{amp}}
	\end{equation}

	Pure dephasing, on the other hand, describes the loss of quantum coherence without energy exchange. This process is governed by the following collapse operators:
	\begin{equation}
		L^{(0)}_z = \sigma^{(0)}_z, \quad \text{with rate } \gamma_{\text{deph}}
	\end{equation}
	\begin{equation}
		L^{(1)}_z = \sigma^{(1)}_z, \quad \text{with rate } \gamma_{\text{deph}}
	\end{equation}

	In addition to the constant Markovian decay, a time-dependent dephasing mechanism is introduced, which is triggered specifically by the presence of a driving pulse. The collapse operator for this process is a collective dephasing operator that acts on both qubits simultaneously:
	\begin{equation}
		L_{\text{pulse}} = \sigma^{(0)}_z + \sigma^{(1)}_z
	\end{equation}
	The rate of this process, $\gamma_{\text{pulse}}(t)$, is not constant but is proportional to the square of the pulse amplitude, $f(t)$:
	\begin{equation}
		\gamma_{\text{pulse}}(t) = \mathcal{G} \times \big(f(t)\big)^2
	\end{equation}
	where $\mathcal{G}$ is the pulse-induced dephasing strength.

	\section{Initial Quantum States}
	The simulations of the two-qubit system are initialized at time $t = 0$ with two distinct quantum states, serving as a basis for comparing the evolution of entanglement and other quantum correlations.

	The first initial state is the maximally entangled Bell state, $| \Phi^+ \rangle$, which represents a state with the highest possible degree of entanglement for a two-qubit system. This state is defined as \cite{Bell_1964}:
	\begin{equation}
		| \Phi^+ \rangle = \frac{1}{\sqrt{2}}(|00\rangle + |11\rangle)
	\end{equation}
	The corresponding initial density matrix, $\rho_{\text{Bell}}(0)$, is a pure state projector:
	\begin{equation}
		\rho_{\text{Bell}}(0) = | \Phi^+ \rangle \langle \Phi^+ |
	\end{equation}

	The second initial state is a fully separable product state, specifically the ground state of both qubits. This state lacks any quantum correlation or entanglement by definition. It is expressed as:
	\begin{equation}
		|00\rangle = |0\rangle \otimes |0\rangle
	\end{equation}
	The initial density matrix for this state, $\rho_{\text{Separable}}(0)$, is also a pure state projector:
	\begin{equation}
		\rho_{\text{Separable}}(0) = |00\rangle \langle 00|
	\end{equation}
	
	\section{Quantum Metrics}
	To characterize the system's state during its time evolution, three key quantum measures are calculated. These measures provide insights into the dynamics of entanglement, total quantum correlations, and the mixedness of the state.
	
	\subsection{Negativity ($\mathcal{N}(\rho)$)}
	Negativity is an entanglement monotone used to quantify entanglement in bipartite systems \cite{Vidal_2002}. For a two-qubit state $\rho$, it is formally defined as the absolute sum of the negative eigenvalues of the partial transpose of the density matrix. It is most conveniently calculated using the trace norm of the partial transpose:
	\begin{equation}
		\mathcal{N}(\rho) = \frac{\Vert \rho^{T_A} \Vert_1 - 1}{2}
	\end{equation}
	where $\rho^{T_A}$ is the partial transpose of $\rho$ with respect to subsystem A, and $\Vert X \Vert_1 = \text{Tr}\sqrt{X^\dagger X}$ is the trace norm. For a maximally entangled two-qubit state, the negativity is $\mathcal{N}(\rho)=0.5$. A value of $\mathcal{N}(\rho) > 0$ indicates entanglement. The maximum possible value for a two-qubit state is $0.5$.
	
	\subsection{Geometric Discord ($D_G(\rho)$)}
	Geometric discord is a measure of total quantum correlations, capturing both entanglement and other non-classical correlations \cite{Ollivier_2001}. It is defined as the minimum Hilbert-Schmidt distance between the given state and the set of classical-quantum states. For a two-qubit state $\rho$, with Bloch vectors $\vec{x}$ and $\vec{y}$ and correlation matrix $T$, the density matrix can be expressed as:
	\begin{equation}
		\begin{aligned}
			\rho =& \frac{1}{4} \Big( I \otimes I + \sum_{i} x_i \sigma_i \otimes I + \sum_{j} y_j I \otimes \sigma_j \\& + \sum_{i,j} T_{ij} \sigma_i \otimes \sigma_j \Big)
		\end{aligned}
	\end{equation}
	where $x_i = \text{Tr}[\rho(\sigma_i \otimes I)]$ and $T_{ij} = \text{Tr}[\rho(\sigma_i \otimes \sigma_j)]$. The geometric discord is calculated as:
	\begin{equation}
		D_G(\rho) = \frac{1}{4} \left( \text{Tr}(K) - \lambda_{\max}(K) \right)
	\end{equation}
	where $K = \vec{x}\vec{x}^T + T^T T$, and $\lambda_{\max}(K)$ is the largest eigenvalue of the matrix $K$.
	
	\subsection{Quantum Memory-Assisted Entropic Uncertainty}
	\label{subsec:qm_eur}
	
	The quantum memory-assisted entropic uncertainty relation (QM-EUR) quantifies the uncertainty in measuring two incompatible observables, such as \(\sigma_x\) and \(\sigma_z\), on qubit \(A\) with qubit \(B\) as quantum memory. For a two-qubit state \(\rho_{AB}\), the QM-EUR is \cite{Berta2010}:
	
	\begin{equation}
		S(X_A | B) + S(Z_A | B) \geq \log_2 \frac{1}{c} + S(A | B),
		\label{eq:qm_eur}
	\end{equation}
	
	where \(S(X_A | B) = \sum_k p_k^x S(\rho_{A|k}^x)\) and \(S(Z_A | B) = \sum_k p_k^z S(\rho_{A|k}^z)\) are conditional entropies, with \(p_k^x = \tr(\rho_{AB} (I_A \otimes \Pi_k^x))\), \(\rho_{A|k}^x = \tr_B[(I_A \otimes \Pi_k^x) \rho_{AB}]/p_k^x\), and similarly for \(\sigma_z\). Here, \(c = 1/2\) for \(\sigma_x\) and \(\sigma_z\), so \(\log_2 (1/c) = 1\), and \(S(A | B) = S(\rho_{AB}) - S(\rho_B)\).
	
	For the Bell state \(|\Phi^+\rangle = (|00\rangle + |11\rangle)/\sqrt{2}\), maximal entanglement yields \(S(X_A | B) \approx 0\), \(S(Z_A | B) \approx 0\), and \(S(A | B) \approx -1\), giving QM-EUR \(\approx 0\). For the separable state \(|00\rangle\), no correlations result in \(S(X_A | B) \approx 1\), \(S(Z_A | B) \approx 1\), and \(S(A | B) \approx 0\), so QM-EUR \(\approx 2\).
	
	In the simulation, QM-EUR is approximated using negativity \(\mathcal{N}(\rho) = (\norm{\rho^{T_A}}_1 - 1)/2\):
	
	\begin{equation}
		U \approx 2 (1 - 2 \mathcal{N}(\rho)),
	\end{equation}
	This simplified linear mapping is introduced as a heuristic relation between
	the quantum-memory entropic uncertainty and the degree of entanglement.
	In our simulations, the initial states were either maximally entangled
	($|\psi^+\rangle$) or separable ($|00\rangle$), and the dynamics remained confined
	to weakly mixed, near-pure subspaces under pulse-induced dephasing.
	In this regime, both the negativity and the conditional entropic terms vary
	almost linearly with the degree of decoherence. We therefore adopt Eq.~(22) as an
	approximate tool to visualize the interplay between entanglement and uncertainty.
	This is not a general identity, but an empirical relation valid for near-pure
	states and weak decoherence conditions.
	ensuring \(U \approx 0\) for the Bell state (\(\mathcal{N} = 0.5\)) and \(U \approx 2\) for the separable state (\(\mathcal{N} = 0\)) \cite{Coles2017}.

	\section{Results and Discussion}
	
	We are examining the dynamics of three significant quantum correlation measures in this collection of graphs: NG, QD, and QM-EUR. Our results show how these quantum characteristics change over time when a driving pulse is applied. There are two initial conditions for the system: the Bell state, which has strong quantum entanglement (\(|\psi^+\rangle\)), and the separable state, which has no entanglement at all (\(|\psi_{00}\rangle\)).

	\begin{figure*}
		\centering
		\includegraphics[width=0.9\linewidth]{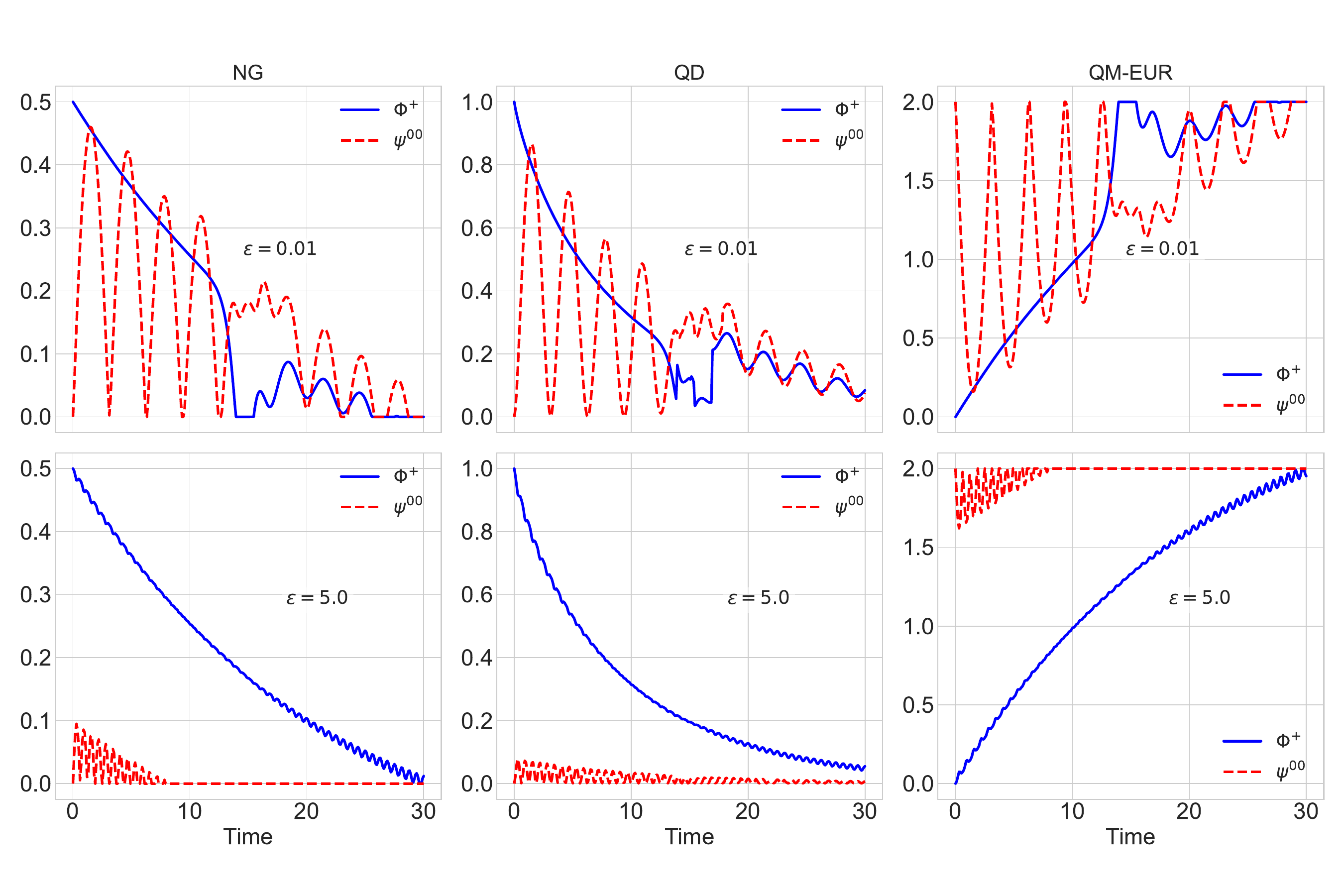}
		\caption{Dynamics of quantum correlations under a driving pulse. The evolution of Negativity, Quantum Discord, and QM-Assisted Entropic Uncertainty is compared for a system starting in either a Bell state ($|\psi^{+}\rangle$) or a separable state ($|\psi^{00}\rangle$). The dynamics are strongly dependent on the qubit energy splitting, with the top row showing the weak-coupling regime ($\varepsilon = 0.01$) and the bottom row showing the strong-coupling regime ($\varepsilon = 5.0$). Simulation parameters: $J_{zz} = J_{xx} = 0.5$, $\gamma_{\text{amp}} = 0.01$, $\gamma_{\text{deph}} = 0.01$, $G = 0.01$. The driving pulse has an amplitude $A_{\text{pulse}} = 1.0$ and is centered at $t=15$.}
		\label{figx1.pdf}
	\end{figure*}

	Fig.~\ref{figx1.pdf} presents a detailed comparison of the time evolution of three quantum correlation measures—NG, QD, and QM-EUR, for a two-qubit system subjected to a driving pulse. The analysis contrasts two distinct initial states: the maximally entangled Bell state ($\ket{\Phi^+}$, solid blue) and the separable state ($\ket{\psi^{00}}$, dashed red), across two regimes of qubit energy splitting: weak coupling ($\varepsilon = 0.01$, top row) and strong coupling ($\varepsilon = 5.0$, bottom row). In the weak-coupling regime, the three measures show noticeable oscillations for both initial states. This behavior comes from the coherent exchange between the driving pulse and the system’s own dynamics. For the Bell state, NG and QD begin at their highest values and then slowly decrease, with a few small revivals triggered by the pulse. The separable state, on the other hand, undergoes stronger oscillations and larger revivals in both NG and QD, showing that the driving pulse can efficiently create short-lived quantum correlations from an initially uncorrelated state. The QM-EUR measure also oscillates, with both states reaching similar values at late times, suggesting that the pulse can dynamically erase the initial-state memory in this regime. A starkly different picture emerges in the strong-coupling regime $\varepsilon = 5.0$. Here, the Bell state maintains a clear advantage: both NG and QD for $\ket{\Phi^+}$ decay smoothly but remain significantly higher than those for $\ket{\psi^{00}}$, which quickly drop to near zero and remain suppressed throughout the evolution. This indicates that strong intrinsic energy splitting protects the initial entanglement of the Bell state but inhibits the generation of correlations from a separable state. The QM-EUR measure for the Bell state shows a monotonic increase, reflecting the persistent quantumness of the system, while for the separable state, QM-EUR saturates rapidly, indicating a lack of further correlation development. These results show the crucial role of both the initial state and the system's energy structure in determining the fate of quantum correlations under external driving. In the weak-coupling regime, the system is highly susceptible to the driving pulse, allowing even separable states to acquire significant, albeit transient, quantum correlations. In contrast, strong coupling stabilizes pre-existing entanglement but suppresses the ability of the drive to generate new correlations from unentangled states. The QM-EUR measure reveals that the driving field can temporarily increase the system’s uncertainty. Over time, this uncertainty settles to a steady level, determined by the balance between the system’s intrinsic energy scales and the initial quantum correlations.\\

	\begin{figure*}
		\centering
		\includegraphics[width=0.9\linewidth]{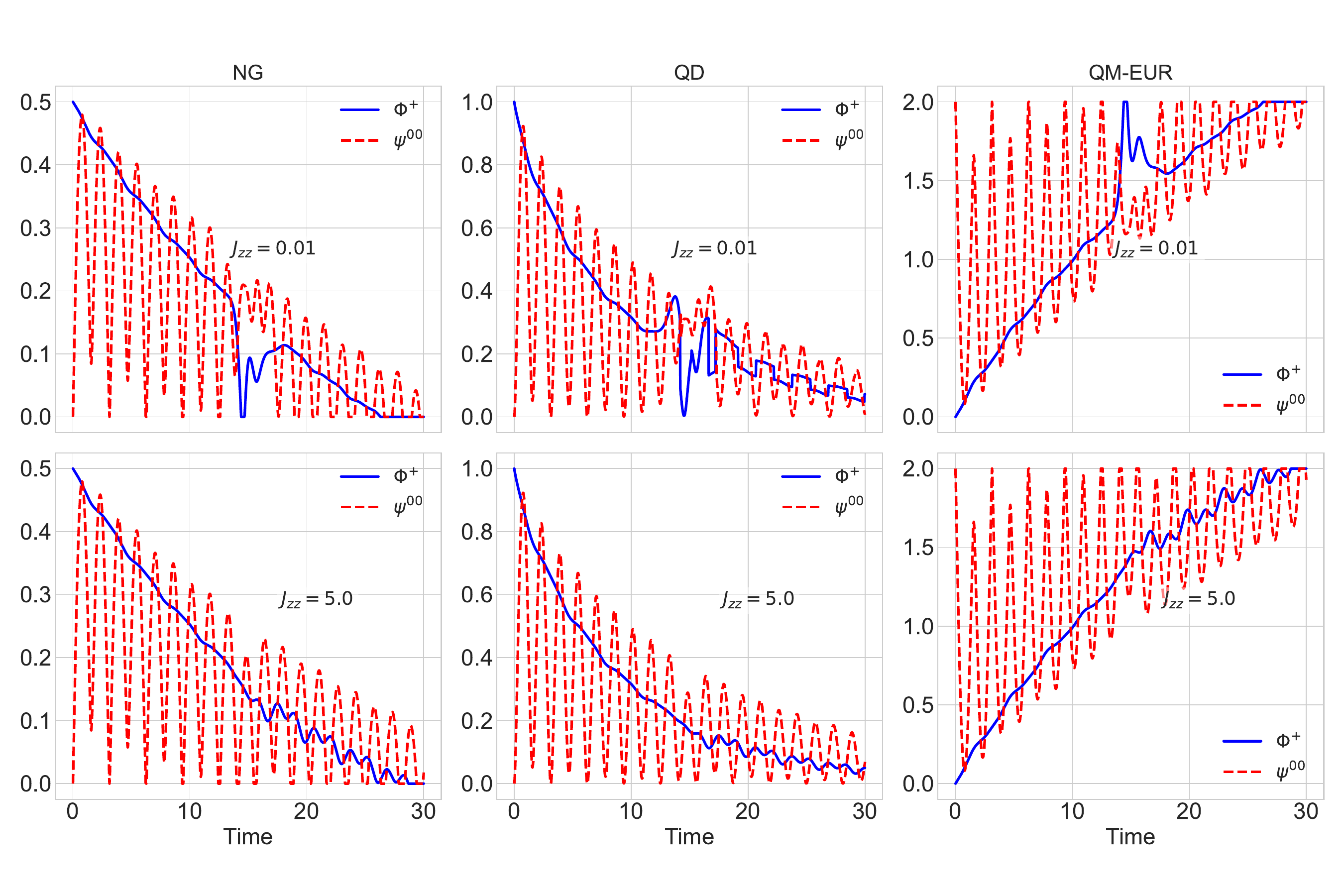}
		\caption{Dynamics of quantum correlations under a driving pulse. The evolution of Negativity (NG), Quantum Discord (QD), and QM-Assisted Entropic Uncertainty (QM-EUR) is compared for a system starting in either a Bell state ($|\psi^{+}\rangle$) or a separable state ($|\psi^{00}\rangle$). The dynamics are shown to be strongly dependent on the ZZ-coupling strength, with the top row corresponding to a weak interaction ($J_{zz} = 0.01$) and the bottom row to a strong interaction ($J_{zz} = 5.0$). Other simulation parameters are fixed: qubit energy splitting $\varepsilon=0.1$, XX-coupling $J_{xx}=1.0$, amplitude damping $\gamma_{\text{amp}}=0.01$, pure dephasing $\gamma_{\text{deph}}=0.01$, and $G=0.01$. The driving pulse has an amplitude $A_{\text{pulse}}=1.0$ and is centered at $t=15$.}
		\label{figx2.pdf}
	\end{figure*}

	Figure~\ref{figx2.pdf} illustrates the evolution of quantum correlations as quantified by Negativity (NG), Quantum Discord (QD), and the Quantum Memory–Based Uncertainty Relation (QM-EUR). The results are presented for two initial configurations: the maximally entangled Bell state $\ket{\psi^+}$ and the separable state $\ket{00}$. The analysis is carried out in two coupling regimes: weak interaction ($J_{zz} = 0.01$, top row) and strong interaction ($J_{zz} = 5.0$, bottom row).
	
	In the weak coupling regime ($J_{zz} = 0.01$), all three correlation measures exhibit oscillatory behavior that depends strongly on the initial state. For the Bell state, the initial values of NG and QD (NG $\approx 0.5$ and QD $\approx 1$) gradually decrease while showing oscillations driven by the external pulse, reflecting the persistence of quantum correlations despite moderate decoherence. Around $t = 15$, corresponding to the pulse application, the Bell state undergoes a sudden perturbation characterized by an abrupt loss of correlations followed by partial regeneration. This transient decay–revival sequence indicates a redistribution of coherence and a non-Markovian response to the external drive. At the same time, QM-EUR exhibits a sharp peak, confirming a temporary increase in quantum uncertainty caused by the pulse. For the separable state $\ket{00}$, the appearance of a pronounced NG peak immediately after the pulse indicates the creation of transient entanglement induced by the external field, although this correlation rapidly vanishes due to dissipation. The strong oscillations and correlation revivals observed in both NG and QD demonstrate the pulse’s effectiveness in generating short-lived quantum correlations even from an initially uncorrelated state. Notably, QD remains nonzero even when NG vanishes, emphasizing the resilience of nonclassical correlations that extend beyond entanglement. The evolution of QM-EUR behaves inversely to that of NG and QD. When the system is maximally entangled, QM-EUR reaches its minimum (QM-EUR = 0), reflecting the suppression of uncertainty by strong quantum correlations. As time evolves, QM-EUR increases monotonically while NG and QD decrease, revealing a redistribution of information and a gradual reduction of correlations. For the separable state, QM-EUR eventually approaches its maximum value (QM-EUR = 2), indicating the absence of correlations and the dominance of uncertainty. These observations suggest that in the weak coupling regime, the applied pulse progressively erases the memory of the initial configuration, highlighting the system’s sensitivity to external driving and the key role of the coupling strength in controlling short-lived quantum correlations.
	
	In the strong coupling regime ($J_{zz} = 5.0$), the behavior of NG, QD, and QM-EUR for the Bell state $\ket{\psi^+}$ shows only minor differences compared to the weak coupling case. The overall level of entanglement and quantum correlations changes little, whereas the separable state $\ket{00}$ remains nearly unaffected. However, the oscillations become more regular and coherent, indicating that strong coupling helps stabilize the quantum dynamics against dissipation and delays the onset of decoherence. Under strong coupling, the qubits are more responsive to the external pulse, producing larger fluctuations in the Bell state. This enhanced sensitivity underlines the crucial role of the interaction term $J_{zz}$, which amplifies correlation effects and strengthens the system’s resilience to external perturbations. Even the separable state displays slightly stronger fluctuations than in the weak coupling regime, suggesting that $J_{zz}$ also influences the stability and dynamical response of unentangled configurations.
	
	Overall, the results demonstrate that the interplay between $J_{zz}$ and the initial state critically determines the evolution of quantum correlations. The coupling parameter $J_{zz}$ acts as a key control factor modulating both the amplitude of fluctuations and the system’s response to external driving, whereas the nature of the initial state governs the rate and robustness of correlation buildup. An initially entangled state tends to stabilize faster and withstand perturbations more effectively, while a separable state remains largely uncorrelated, generating only short-lived correlations. These findings highlight the interdependence between interaction strength and initial configuration in the preservation, generation, and control of quantum resources, underscoring their central importance for quantum information and technological applications.

	\begin{figure*}
		\centering
		\includegraphics[width=0.9\linewidth]{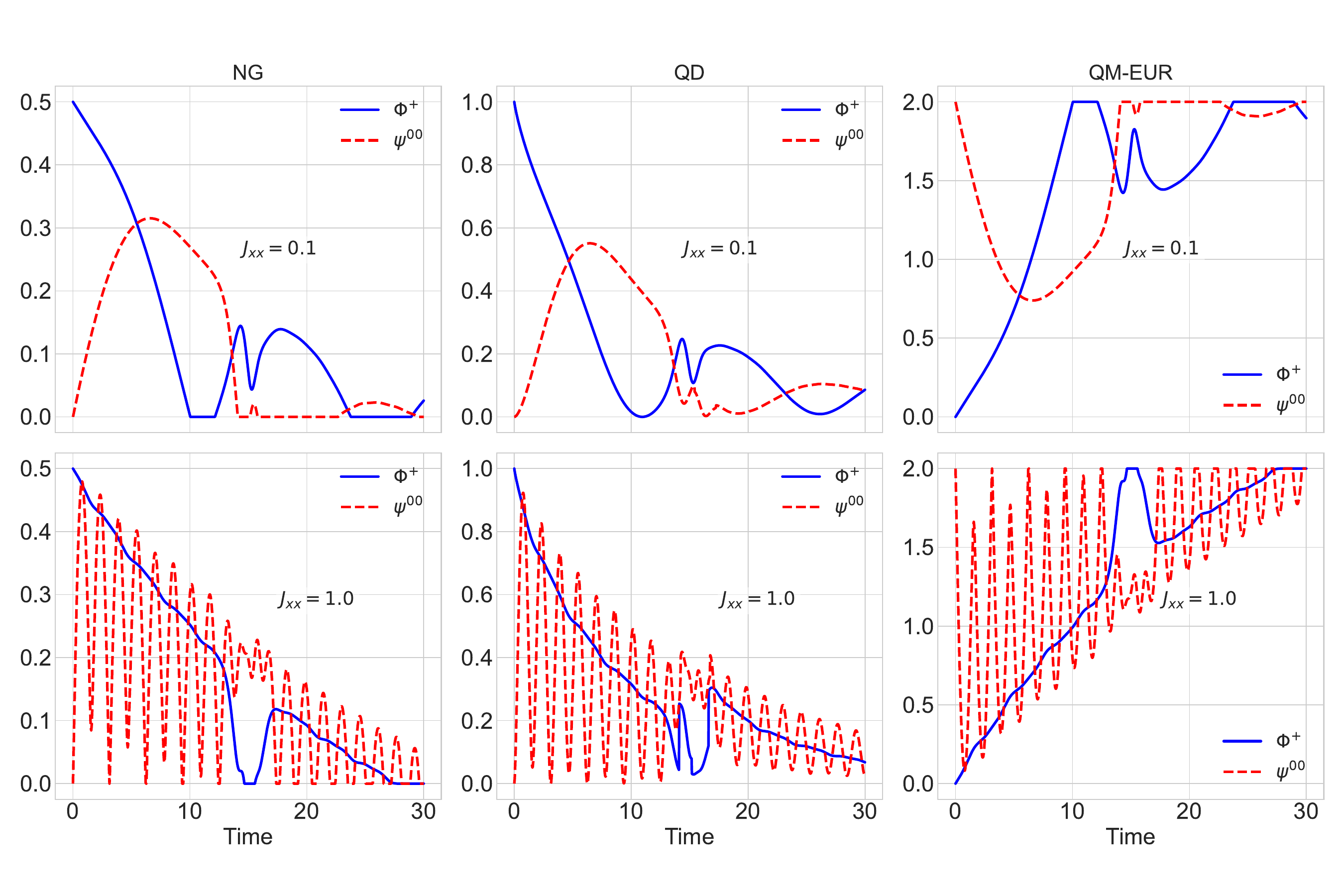}
		\caption{Dynamics of quantum correlations under a driving pulse. The evolution of Negativity (NG), Quantum Discord (QD), and QM-Assisted Entropic Uncertainty (QM-EUR) is compared for a system starting in either a Bell state ($|\psi^{+}\rangle$) or a separable state ($|\psi^{00}\rangle$). The dynamics are shown to be dependent on the XX-coupling strength, with the top row corresponding to a weak coupling ($J_{xx} = 0.1$) and the bottom row to a stronger coupling ($J_{xx} = 1.0$). Other simulation parameters are fixed: qubit energy splitting $\epsilon=0.1$, ZZ-coupling $J_{zz}=1.0$, amplitude damping $\gamma_{\text{amp}}=0.01$, pure dephasing $\gamma_{\text{deph}}=0.01$, and $G=0.01$. The driving pulse has an amplitude $A_{\text{pulse}}=1.0$ and is centered at $t=15$.}
		\label{figx3.pdf}
	\end{figure*}

	Figure~\ref{figx3.pdf} shows the temporal evolution of Negativity (NG), Quantum Discord (QD), and the Quantum-Memory-Assisted Entropic Uncertainty (QM-EUR) under the influence of an excitation pulse for two values of the XX coupling: $J_{xx}=0.1$ and $J_{xx}=1.0$. The simulations were performed with the following fixed parameters: qubit energy separation $\varepsilon = 0.1$, ZZ coupling $J_{zz} = 1.0$, amplitude damping $\gamma_{\rm amp} = 0.01$, pure dephasing $\gamma_{\rm deph} = 0.01$, and $G = 0.01$. The applied pulse has an amplitude $A_{\rm pulse} = 1.0$ and is centered around $t = 15$. These conditions allow for a clear assessment of the effect of XX coupling and the pulse on the dynamics of correlations and quantum memory.
	
	In the weak coupling regime ($J_{xx}=0.1$), Negativity for the Bell state $|\psi^+\rangle$ starts at its maximum value ($NG \simeq 0.5$) and gradually decreases, displaying small oscillations after the pulse. This behavior indicates a partial loss of entanglement, with the initially strongly correlated state becoming weakly entangled due to both the weak interaction and the external perturbation. For the separable state $|00\rangle$, although there is no initial entanglement, NG exhibits small oscillations reaching approximately 0.31 around $t = 7$, showing that weak coupling can still temporarily generate quantum correlations. Quantum Discord highlights the resilience of correlations in this regime. For the Bell state, QD remains high with only minor oscillations, suggesting that some nonclassical correlations persist even when entanglement measured by NG decreases. For the separable state, QD reaches a maximum of about 0.59 at $t = 0.5$ but almost completely vanishes afterwards, emphasizing that external perturbations can significantly affect initially weak correlations. The QM-EUR exhibits a contrasting behavior depending on the initial state. For the Bell state, uncertainty starts low but rapidly rises to values close to 2, with pronounced oscillations reflecting the entanglement dynamics and its influence on quantum memory. The separable state begins with high uncertainty, which temporarily decreases before stabilizing around oscillations near 2, illustrating the limited ability of the system to reduce uncertainty in the absence of entanglement. At $t = 15$, the pulse abruptly modifies the oscillations for both states, revealing the system's sensitivity to external perturbations.
	
	In the strong coupling regime ($J_{xx}=1.0$), the dynamics of quantum correlations are significantly altered. For the separable state, initially unentangled, NG and QD show larger amplitude and higher-frequency oscillations, indicating that strong coupling can dynamically generate correlations and trigger non-Markovian feedback cycles. The pulse applied at $t = 15$ nearly suppresses the correlations, which reappear as rapid, low-amplitude oscillations, illustrating the system's sensitivity to perturbations and its ability to partially regenerate correlations. The QM-EUR for the Bell state continues to increase after the pulse, accompanied by rapid oscillations, signaling a competition between coherence recovery and dissipation. For the separable state, the oscillations become irregular and stabilize at a higher asymptotic value, reflecting the persistence of mostly classical uncertainty in the absence of significant quantum resources. These results demonstrate that XX coupling plays a crucial role in the formation, stability, and lifetime of quantum correlations. Weak coupling partially preserves the initial coherence and produces only a small amount of new correlations, whereas strong coupling generates complex oscillations characteristic of non-Markovian behavior and reveals quantum memory effects. The combined analysis of NG, QD, and QM-EUR confirms that quantum memory is an essential resource to maintain and control the robustness of correlations, and that the external pulse can significantly modulate these correlations depending on the coupling strength and the initial state of the system.
	\\
	
	\begin{figure*}
		\centering
		\includegraphics[width=0.9\linewidth]{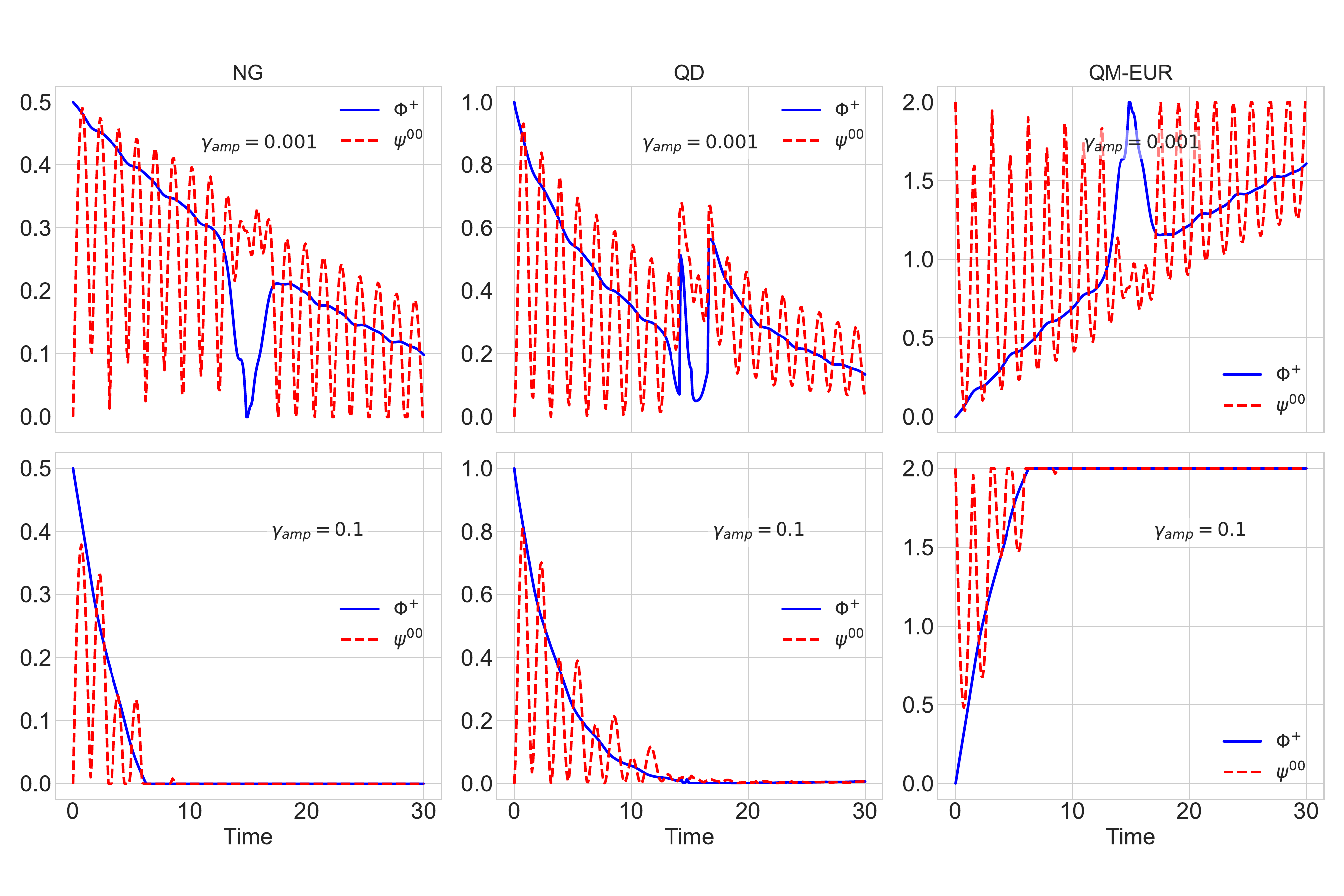}
		\caption{Dynamics of quantum correlations under a driving pulse. The evolution of Negativity (NG), Quantum Discord (QD), and QM-Assisted Entropic Uncertainty (QM-EUR) is compared for a system starting in either a Bell state ($|\psi^{+}\rangle$) or a separable state ($|\psi^{00}\rangle$). The dynamics are shown for two different amplitude damping rates, with the top row corresponding to a weak damping ($\gamma_{\text{amp}} = 0.01$) and the bottom row to a stronger damping ($\gamma_{\text{amp}} = 0.1$). Other simulation parameters are fixed: qubit energy splitting $\varepsilon=0.1$, couplings $J_{zz}=J_{xx}=1.0$, pure dephasing $\gamma_{\text{deph}}=0.01$, and $G=0.01$. The driving pulse has an amplitude $A_{\text{pulse}}=1.0$ and is centered at $t=15$.}
		\label{figx4.pdf}
	\end{figure*}
	
	Fig.~\ref{figx4.pdf} shows the comparative dynamics of three quantum correlation measures, namely Negativity (NG), Quantum Discord (QD), and Quantum-Memory Assisted Entropic Uncertainty (QM-EUR), for a two-qubit system subjected to a control pulse in the presence of noise. The analysis considers two initial states, the maximally entangled Bell state $\lvert \psi^+ \rangle$ and the separable state $\lvert \psi_{00} \rangle$, under two distinct damping regimes, weak ($\gamma_{\text{amp}} = 0.01$) and strong ($\gamma_{\text{amp}} = 0.1$), which are represented respectively by the first and second rows of the figure.
	
	In the weak damping regime ($\gamma_{\text{amp}} = 0.01$), NG and QD exhibit similar evolution for the Bell state, although QD reaches a higher amplitude. At $t = 0$, the initial values are (NG $\approx 0.5$) and (QD $\approx 1$), reflecting a high level of quantum correlations. For the separable state, small damped oscillations occur for $t \neq 0$, reflecting temporary fluctuations in correlations caused by weak damping. When the pulse is applied at $t = 15$, the Bell state curves, including QM-EUR, experience a sharp drop, indicating a brief reduction in quantum correlations as energy is transferred from the external field. The pulse acts as a resonant perturbation, momentarily disturbing entanglement and redistributing correlations within the system. Following the pulse, QM-EUR gradually recovers, showing partial restoration of correlations, while NG and QD decrease simultaneously. By contrast, for the separable state, QM-EUR remains high and nearly constant, consistent with the initially low entanglement and the relative stability of correlations under external perturbations. These results underline that initially entangled states are more sensitive to external disturbances, whereas separable states are comparatively resilient.
	
	In the strong damping regime ($\gamma_{\text{amp}} = 0.1$), two distinct behaviors emerge depending on the purity of the initial state. For the Bell state, QD reaches its maximum value as $t \rightarrow 0$, confirming a high initial level of correlation. For $t > 0$, NG decreases monotonically and vanishes at a critical time ($t_c \approx 6$), at which point the density matrix becomes separable. QD, being more resilient, indicates that some correlations persist even in the absence of entanglement. For the separable state, oscillations are weak and rapidly damped, and both NG and QD tend toward zero as $t \rightarrow \infty$. QM-EUR reaches a maximum at $t \approx 6$, reflecting the maximum measurement uncertainty resulting from the absence of quantum memory and the complete degradation of correlations. These findings emphasize that precise control of damping is crucial for preserving quantum correlations, a key requirement for quantum information and communication protocols.
	
	Comparing the two damping regimes shows that the number of oscillations and how long correlations last depend heavily on $\gamma_{\text{amp}}$. With weak damping, correlations persist for a longer time and the oscillations remain fairly regular, suggesting mostly coherent and reversible dynamics. On the other hand, strong damping causes correlations to drop quickly, even when a pulse is applied. This figure highlights how important damping is for sustaining quantum correlations and suggests that the stability of quantum effects relies on a careful balance between external driving and decoherence.
	
	\begin{figure*}
		\centering
		\includegraphics[width=0.9\linewidth]{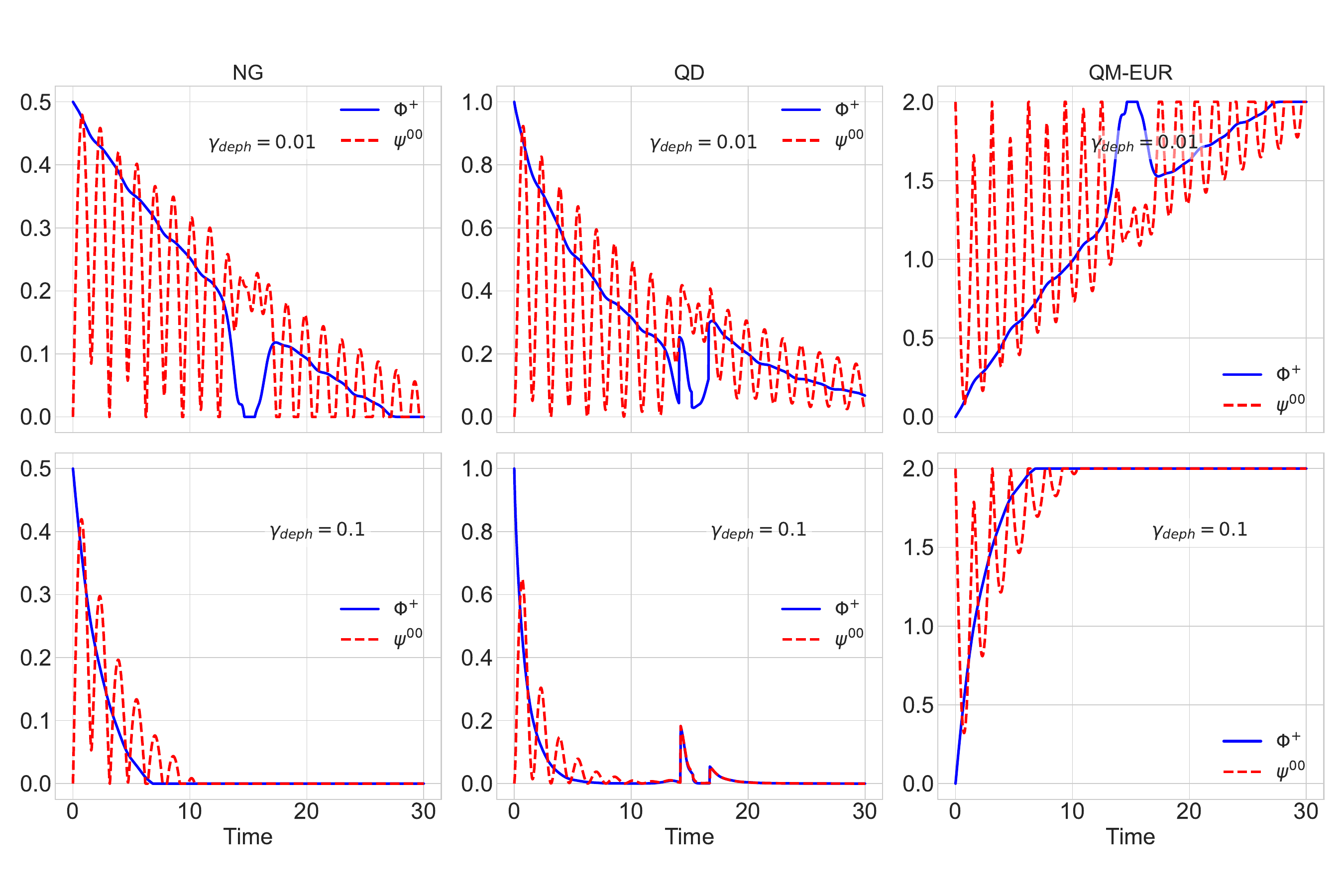}
		\caption{Dynamics of quantum correlations under a driving pulse. The evolution of Negativity (NG), Quantum Discord (QD), and QM-Assisted Entropic Uncertainty (QM-EUR) is compared for a system starting in either a Bell state ($|\psi^{+}\rangle$) or a separable state ($|\psi^{00}\rangle$). The dynamics are shown for two different pure dephasing rates, with the top row corresponding to weak dephasing ($\gamma_{\text{deph}} = 0.01$) and the bottom row to stronger dephasing ($\gamma_{\text{deph}} = 0.1$). Other simulation parameters are fixed: qubit energy splitting $\varepsilon=0.1$, couplings $J_{zz}=J_{xx}=1.0$, amplitude damping $\gamma_{\text{amp}}=0.01$, and $G=0.01$. The driving pulse has an amplitude $A_{\text{pulse}}=1.0$ and is centered at $t=15$.}
		\label{figx5.pdf}
	\end{figure*}

	\begin{figure*}
		\centering
		\includegraphics[width=0.9\linewidth]{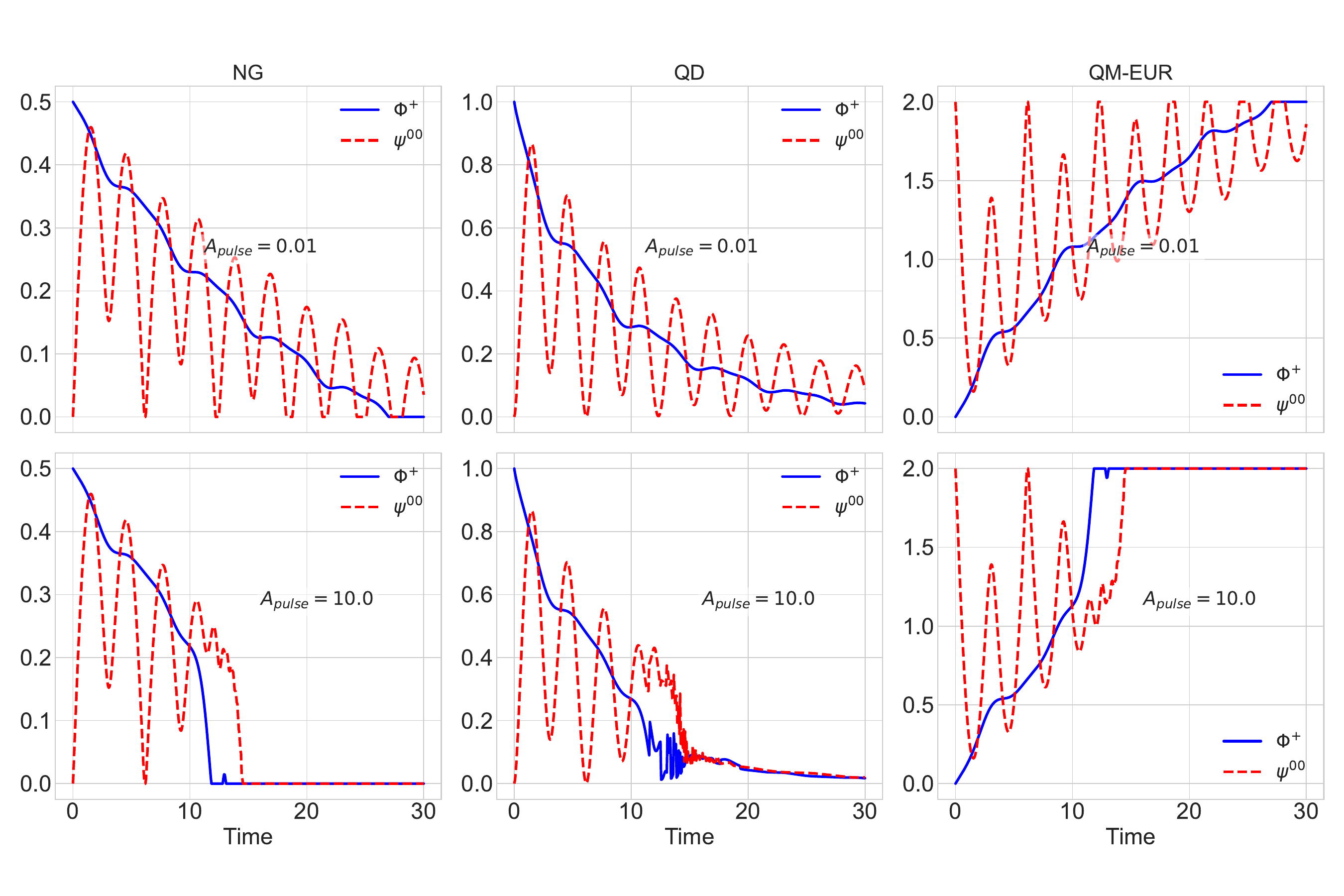}
		\caption{Dynamics of quantum correlations under a driving pulse. The evolution of Negativity (NG), Quantum Discord (QD), and QM-Assisted Entropic Uncertainty (QM-EUR) is compared for a system starting in either a Bell state ($|\psi^{+}\rangle$) or a separable state ($|\psi^{00}\rangle$). The dynamics are shown for two different driving pulse amplitudes, with the top row corresponding to a weak pulse ($A_{\text{pulse}} = 0.01$) and the bottom row to a strong pulse ($A_{\text{pulse}} = 10.0$). Other simulation parameters are fixed: qubit energy splitting $\varepsilon=0.1$, couplings $J_{zz}=J_{xx}=0.5$, amplitude damping $\gamma_{\text{amp}}=0.01$, pure dephasing $\gamma_{\text{deph}}=0.01$, and $G=0.01$. The driving pulse is centered at $t=15$.}
		\label{figx6.pdf}
	\end{figure*}

	\begin{figure*}
		\centering
		\includegraphics[width=0.9\linewidth]{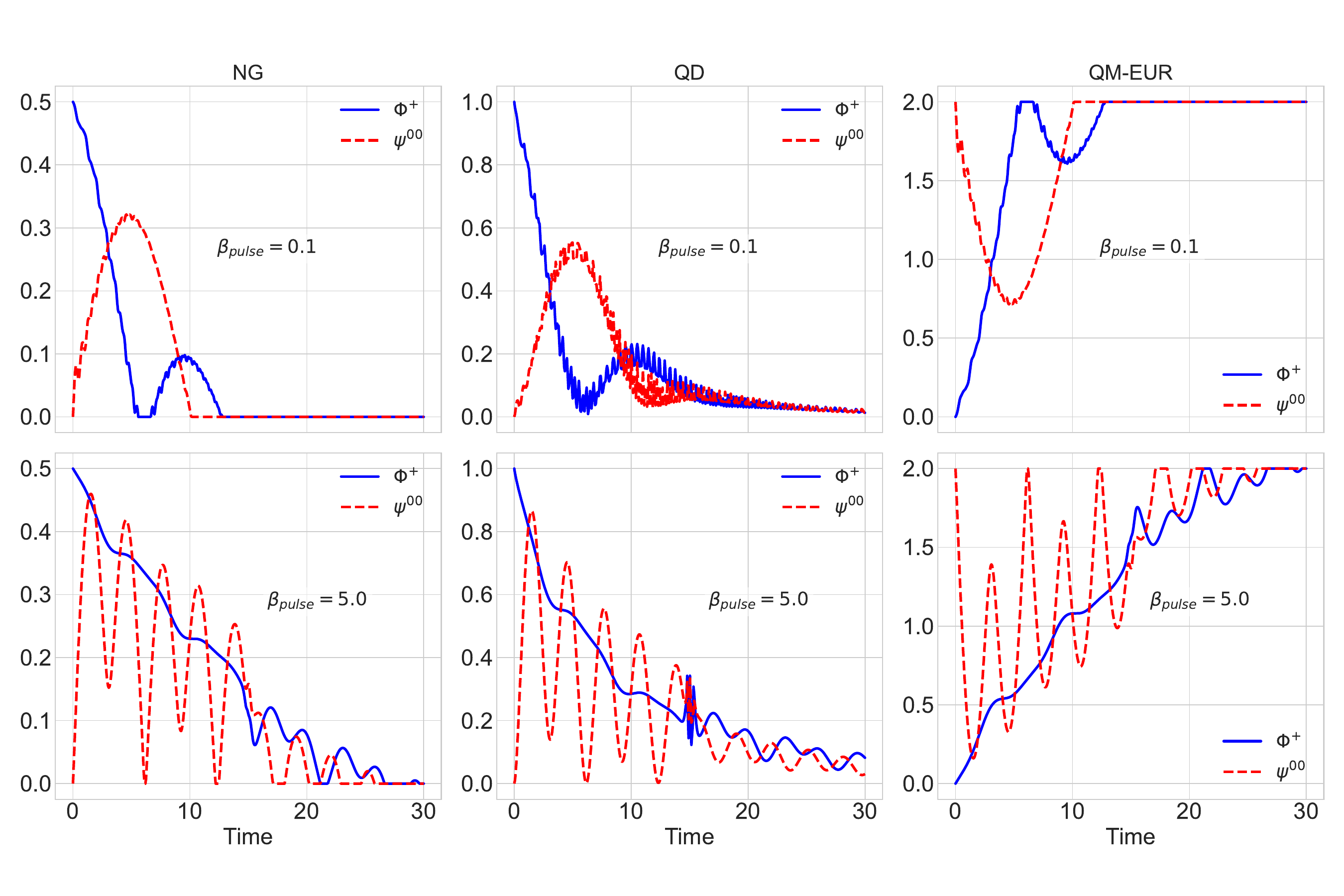}
		\caption{Dynamics of quantum correlations under a driving pulse. The evolution of Negativity (NG), Quantum Discord (QD), and QM-Assisted Entropic Uncertainty (QM-EUR) is compared for a system starting in either a Bell state ($|\psi^{+}\rangle$) or a separable state ($|\psi^{00}\rangle$). The dynamics are shown for two different driving pulse widths, with the top row corresponding to a broad, slow pulse ($\beta_{\text{pulse}} = 0.1$) and the bottom row to a narrow, sharp pulse ($\beta_{\text{pulse}} = 5.0$). Other simulation parameters are fixed: qubit energy splitting $\varepsilon=0.1$, couplings $J_{zz}=J_{xx}=0.5$, pulse amplitude $A_{\text{pulse}}=5.0$, amplitude damping $\gamma_{\text{amp}}=0.01$, pure dephasing $\gamma_{\text{deph}}=0.01$, and $G=0.01$. The driving pulse is centered at $t=15$.}
		\label{figx7.pdf}
	\end{figure*}

	Fig.~\ref{figx5.pdf} shows the dynamics of quantum correlations under the effect of a pulse applied to a two-qubit system subjected to pure dephasing. Three indicators are analyzed: Negativity (NG), which measures the degree of bipartite entanglement; Quantum Discord (QD), which quantifies non-classical quantum correlations beyond entanglement; and QM-assisted Entropic Uncertainty (QM-EUR), which relates quantum correlations to the reduction of entropic uncertainty in local measurements. The curves compare two initial states: the Bell state $\Phi^+$ and the separable state $\psi^{(00)}$, each studied under two pure dephasing strengths, weak ($\gamma_{\rm deph} = 0.01$) and strong ($\gamma_{\rm deph} = 0.1$).
	
	In the weak dephasing regime ($\gamma_{\rm deph} = 0.01$), the dynamics of quantum correlations show a clear contrast between the two initial states. For the Bell state $\Phi^+$, the Negativity starts with a high value, close to 0.5, corresponding to almost elevated entanglement. This entanglement gradually decreases to about 0.2 around $t = 13$. A sharp drop occurs at $t = 15$, coinciding with the pulse application, which leads to the near-total loss of quantum resources and a sudden decorrelation between the qubits. Afterwards, the curve stabilizes at a low value without noticeable oscillations. This behavior reflects a destructive effect of the pulse on the initially entangled state, indicating that the external interaction tends to degrade entanglement rather than regenerate it. In contrast, for the separable state $|\psi^{00}\rangle$, the Negativity is initially almost zero, indicating the absence of entanglement at $t = 0$. The system exhibits clear oscillatory behavior over time, showing the repeated rise and fall of quantum correlations between the qubits. These oscillations gradually lose strength, indicating a slow reduction of coherence due to dephasing. At $t = 15$, when the pulse is applied, the system shows remarkable stability: unlike the Bell state $|\psi^+\rangle$, it does not suddenly lose its correlations. Instead, they continue as weaker oscillations around a non-zero average, highlighting ongoing interaction between the qubits even under external perturbation. Quantum Discord (QD) behaves similarly but proves more resilient, decaying more slowly and staying above the Negativity. This indicates that QD captures a broader range of quantum correlations and is less affected by dephasing. For the Bell state ($|\psi^+\rangle$), the QM-assisted entropic uncertainty decreases only slightly over time, reflecting the weak correlations present. Overall uncertainty in the system drops minimally, suggesting that under weak dephasing, the Bell state’s entanglement contributes little to reducing quantum uncertainty. In contrast, the separable state ($|\psi^{00}\rangle$) maintains high uncertainty throughout, due to the lack of entanglement and quantum correlations.
	
	In the strong dephasing regime ($\gamma_{\rm deph} = 0.1$), decoherence dominates the system’s behavior, leading to a rapid attenuation of quantum correlations. Negativity falls nearly to zero before $t = 10$, indicating the almost complete disappearance of entanglement for both states. Quantum Discord, which is generally more resilient, gradually decreases, but at $t = 15$, it shows a small rise ($QD \approx 0.2$) caused by the applied pulse. At the same time, the QM-assisted entropic uncertainty stops oscillating, signaling that quantum correlations have been fully suppressed by dephasing. Overall, this figure clearly highlights the competing roles of the pulse and dephasing in the dynamics of quantum correlations. When dephasing is weak, the pulse can create, amplify, and modulate correlations between qubits, even for an initially separable state. In contrast, when dephasing is strong, the loss of quantum information dominates the dynamics, leading to a rapid extinction of all forms of correlation. Among the three measures, Negativity vanishes first, confirming the fragility of entanglement; Quantum Discord persists longer, demonstrating its robustness against decoherence; and QM-EUR reflects these effects in terms of global entropic evolution. These results underscore the hierarchy of stability among different forms of quantum correlations in a system subjected to pure dephasing.\\
	
	The dynamics of quantum correlations under a driving pulse are illustrated in Fig.~\ref{figx6.pdf}, comparing three measures: Negativity (NG), Quantum Discord (QD), and the QM-assisted Entropic Uncertainty Relation (QM-EUR). The system is initialized either in a maximally entangled Bell state \(|\psi^{+}\rangle\) or in a separable state \(|\psi_{00}\rangle\). Two excitation regimes are considered, a weak pulse \((A_{\text{pulse}} = 0.01)\) and a strong pulse \((A_{\text{pulse}} = 10.0)\). The other simulation parameters are fixed as \(\varepsilon = 0.1\), \(J_{zz} = J_{xx} = 0.5\), \(\gamma_{\text{amp}} = 0.01\), \(\gamma_{\text{deph}} = 0.01\), and \(G = 0.01\). The driving pulse is centered around \(t = 15\).  
	
	In the weak-pulse regime, both Negativity and Quantum Discord exhibit similar behavior, gradually decreasing under the effect of decoherence, with QD systematically remaining higher than NG. The correlations arising from the entangled state \(|\psi^{+}\rangle\) vanish more rapidly than those from the separable state \(|\psi_{00}\rangle\), which displays small damped oscillations over time. The QM-EUR evolves differently, showing a gradual increase with small fluctuations, revealing a higher sensitivity of the entropic uncertainty to the driving field. The limited energy of the weak pulse does not generate significant oscillations, indicating moderate robustness of entanglement and enhanced responsiveness of the entropic measure.  For the state \(|\psi^{+}\rangle\), both NG and QD start at their maximal values and then decay monotonically, with slight revivals attributed to the transient action of the pulse. These quantities reach almost zero instantaneously, indicating a rapid decay of entanglement and quantum correlations due to the combined effects of decoherence and pulse-induced dynamics. In contrast, the state \(|\psi_{00}\rangle\) exhibits a much more oscillatory dynamics, characterized by recurrent peaks in NG and QD. This behavior shows that the driving pulse can temporarily generate quantum correlations even from an initially uncorrelated state. The QM-EUR also displays oscillations synchronized with NG and QD before converging, for both initial states, to a value of about \(2\), suggesting that the pulse tends to homogenize the entropic dynamics and erase the memory of the initial state.  
	
	Under stronger excitation, the dynamics become more contrasted and more sensitive to the initial state structure. For the separable state \(|\psi_{00}\rangle\), Negativity evolves oscillatory while maintaining relatively high values, indicating the repeated creation of transient entanglement. In contrast, for the entangled state \(|\psi^{+}\rangle\), NG rapidly decays and vanishes around \(t \approx 12\), revealing an almost instantaneous disentanglement of the initial Bell state. This behavior highlights that a strong-amplitude pulse destroys preexisting entanglement while favoring the generation of new dynamical correlations from an initially separable state.  Quantum Discord exhibits a more complex evolution. For \(|\psi^{+}\rangle\), it shows small random fluctuations around \(t \approx 15\), reflecting a reduced sensitivity to strong excitations. Although entanglement is destroyed, part of the nonclassical quantum correlations persists but in an unstable and noisy form. Conversely, for \(|\psi_{00}\rangle\), QD remains significantly higher, indicating that the intense driving field effectively induces quantum correlations from a classically correlated state.  
	
	Regarding the QM-assisted Entropic Uncertainty Relation, the measure associated with \(|\psi^{+}\rangle\) increases monotonically and stabilizes around \(\mathrm{QM\text{-}EUR} \approx 2\) near \(t \approx 12\). This convergence indicates a shift toward dynamics governed mainly by classical effects. For the state \(|\psi_{00}\rangle\), QM-EUR grows more slowly and saturates near \(t \approx 15\), revealing the continued development of quantum correlations in this strong-excitation regime.
	
	The results demonstrate the significant influence of pulse amplitude on the manipulation and generation of quantum correlations. Weak pulses tend to partially maintain preexisting correlations without inducing substantial new entanglement, whereas strong pulses suppress initial entanglement while promoting the formation of dynamical correlations from separable states. These observations highlight the interplay between decoherence and external driving and underscore the potential for precise control over both the type and duration of quantum correlations through appropriately tailored pulse parameters.

	Fig.~\ref{figx7.pdf} illustrates the dynamics of quantum correlations in a two-qubit system subjected to an external pulse. Three quantifiers are studied: Negativity (NG), Quantum Discord (QD), and Measurement-Assisted Entropic Uncertainty (QM-EUR). The comparison is performed for two contrasting initial states, the maximally entangled Bell state \((|\psi^+\rangle)\) and the separable state \((|\psi^{00}\rangle)\), as well as for two pulse widths: a wide and slow pulse \((\beta_{\text{pulse}}=0.1)\) (first row) and a narrow and fast pulse \((\beta_{\text{pulse}}=5.0)\) (second row). The other simulation parameters are fixed at \(\varepsilon = 0.1\), \(J_{zz} = J_{xx} = 0.5\), \(A_{\text{pulse}}=5.0\), \(\gamma_{\text{amp}} = \gamma_{\text{deph}} = 0.01\), and \(G=0.01\).  
	
	The evolution of Negativity strongly depends on the initial state. For the Bell state, NG starts at its maximum value (approximately 0.5) and gradually decreases due to decoherence, showing oscillations caused by the interplay between the pulse and inter-qubit coupling. These oscillations indicate transient coherence revivals, suggesting that coherent energy exchange can delay the collapse of quantum correlations. For the separable state, Negativity initially zero exhibits a transient peak, demonstrating the dynamic generation of entanglement. The timing and magnitude of this peak provide insight into how pulse parameters influence the onset of correlations in initially uncorrelated systems. NG decays more slowly under a fast pulse, whereas slow pulses lead to a rapid disappearance of entanglement, highlighting the sensitivity of correlations to the temporal structure and amplitude of the driving field.  
	
	Quantum Discord exhibits greater robustness than Negativity, reflecting its sensitivity to a wider range of quantum correlations beyond entanglement. For the Bell state, QD gradually decreases and vanishes around $t \approx 15$. For the separable state, QD rapidly becomes nonzero, reaching a maximum of approximately 0.59 at $t = 6$ before decaying toward $t \approx 10$. Persistent oscillations in QD under fast pulses indicate that certain nonclassical correlations can survive even when entanglement is largely suppressed. The distinct response of QD to different pulse parameters also shows that specific correlations can be selectively enhanced without maintaining entanglement, which may be advantageous for quantum control protocols.

	The Measurement-Assisted Entropic Uncertainty (QM-EUR) generally increases over time, reflecting the progressive loss of information due to decoherence. With a fast pulse, the increase of uncertainty is slower and marked by oscillations, indicating transient correlation recovery, while a slow pulse leads to a nearly linear increase, reflecting accelerated correlation loss. This behavior suggests that entropic measures are particularly sensitive to the competition between external driving and environmental dissipation, providing complementary information to NG and QD. At longer times, differences between initial states diminish, with all trajectories converging toward a regime dominated by maximal uncertainty imposed by decoherence, highlighting that environmental effects ultimately govern long-term stability.  
	
	Overall, Fig.~\ref{figx7.pdf} emphasizes that pulse width, pulse speed, and the initial state are critical parameters for controlling the persistence and generation of quantum correlations. Short, intense pulses allow temporary generation and maintenance of correlations from separable states, whereas long, weak pulses favor rapid dissipation of preexisting correlations. These findings demonstrate the strategic role of pulse engineering in optimizing correlation generation and preservation for practical quantum information applications and provide guidance on synchronizing pulses with decoherence times to maximize quantum control efficiency.

\begin{figure*}
		\centering
		\includegraphics[width=0.9\linewidth]{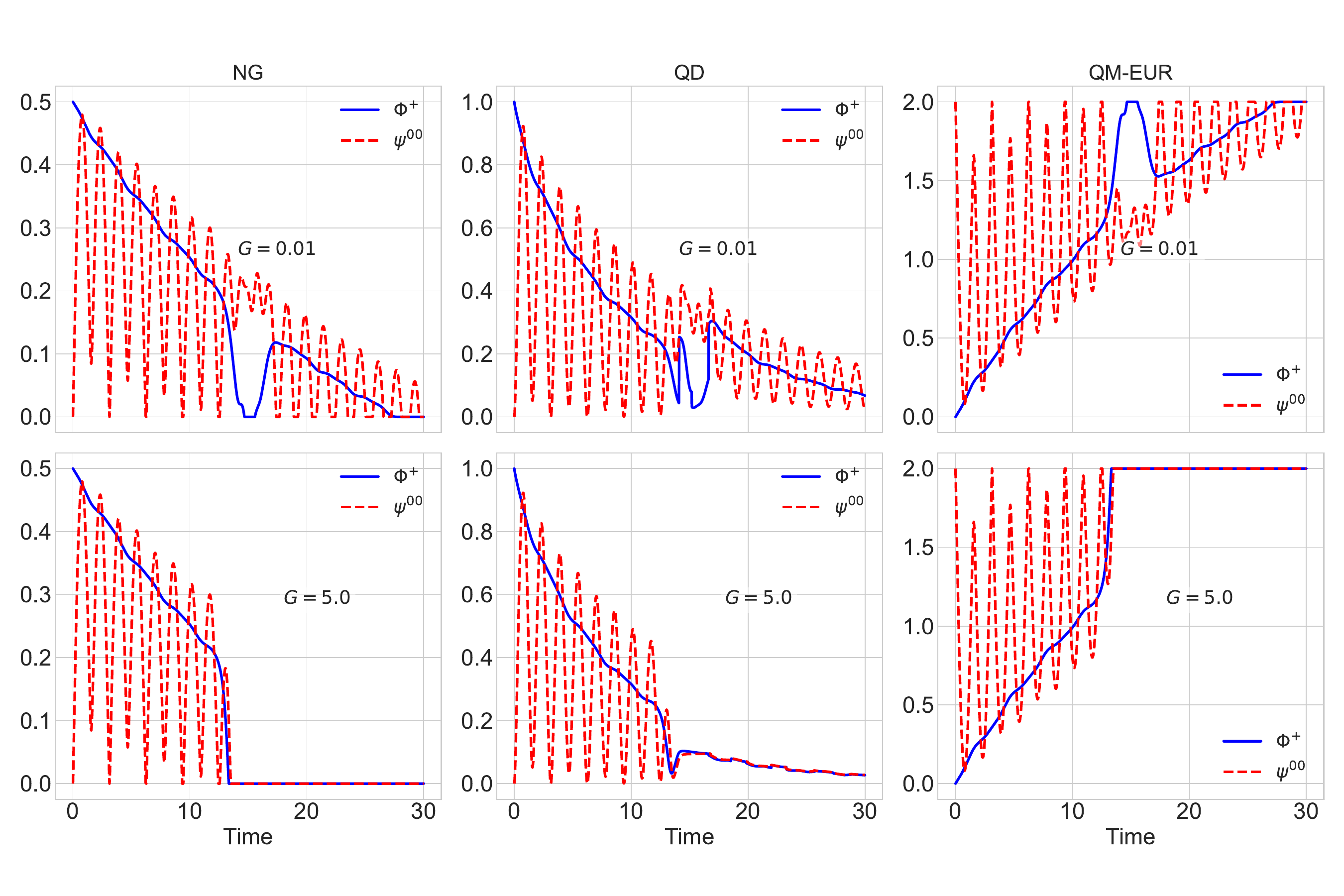}
		\caption{Dynamics of quantum correlations under a driving pulse. The evolution of Negativity (NG), Quantum Discord (QD), and QM-Assisted Entropic Uncertainty (QM-EUR) is compared for a system starting in either a Bell state ($|\psi^{+}\rangle$) or a separable state ($|\psi^{00}\rangle$). The dynamics are shown for two different strengths of pulse-induced dephasing, with the top row corresponding to a weak effect ($G = 0.01$) and the bottom row to a strong effect ($G = 5.0$). Other simulation parameters are fixed: qubit energy splitting $\varepsilon=0.1$, couplings $J_{zz}=j_{xx}=J.0$, amplitude damping $\gamma_{\text{amp}}=0.01$, and pure dephasing $\gamma_{\text{deph}}=0.01$. The driving pulse has an amplitude $A_{\text{pulse}}=1.0$ and is centered at $t=15$.}
		\label{figx8.pdf}
	\end{figure*}

	Figure~\ref{figx8.pdf} illustrates the temporal evolution of quantum correlations, namely Negativity (NG), Quantum Discord (QD), and Quantum-Memory-Assisted Entropic Uncertainty (QM-EUR), in a bipartite system of two qubits subjected to an external pulse. Two dephasing regimes induced by the pulse are considered, a weak regime ($G=0.01$) and a strong regime ($G=5.0$).
	
	For an initially separable state ($|00\rangle$), NG and QD exhibit rapid and sustained oscillations before the pulse ($t<15$), particularly pronounced in the weak dephasing regime. These oscillations reflect the dynamic generation of quantum correlations by the external field, promoting coherent information exchange between the qubits and enabling the temporary emergence of entanglement from an initially uncorrelated state. At $t=15$, the application of the pulse significantly alters this dynamics. In the weak regime, a transient amplification of oscillations is observed, indicating a temporary enhancement of quantum correlations. In contrast, in the strong regime, the pulse disrupts the remaining coherence, causing a rapid decay of NG and QD, signaling increased dissipation of quantum information. The maximum amplitudes, reaching approximately 0.5 for NG and 1 for QD, highlight the robustness of correlations generated under moderate noise. For $G=5.0$, oscillations become strongly damped after the pulse, reflecting accelerated decoherence and phase randomization due to the environment. Remarkably, the QM-EUR remains high throughout the evolution, even after the pulse, indicating the persistence of quantum memory and the presence of intrinsic uncertainty characteristic of the quantum nature of the system.
	
	When the system is initially prepared in a Bell state, the dynamics of correlations show a different behavior. Before the pulse, NG and QD decay rapidly and monotonically, without significant oscillations, reflecting a progressive loss of coherence and the suppression of non-classical correlations by dephasing. Under strong dephasing ($G=5.0$), NG vanishes around $t \approx 13.5$, signaling the complete disappearance of bipartite entanglement. After applying the pulse at $t=15$, no significant regeneration of correlations is observed. NG and QD remain close to zero, indicating that the system, already dominated by noise, cannot recover a coherent regime. Here, the pulse acts as a dissipative agent, reinforcing coherence loss and confirming the irreversible nature of decoherence in this regime. Nevertheless, the QM-EUR slightly increases after the pulse, reflecting a redistribution of quantum resources. Although entanglement is destroyed, the entropic uncertainty becomes the dominant form of non-classicality, indicating residual coherence stored in the system's quantum memory. This analysis highlights the destructive role of dephasing on preexisting entanglement while showing that, even under strong dissipation, the quantum nature of the system does not completely vanish. It manifests in subtler forms, such as increased entropic uncertainty or partial retention of quantum information in the system's memory.
	
	Overall, these results demonstrate the decisive influence of the initial state, dephasing, and applied pulse on the dynamics of quantum correlations. Weak coupling to the environment favors the preservation of coherence and allows oscillatory information exchange, which the pulse can temporarily enhance. Conversely, strong dephasing leads to rapid decorrelation and nearly complete loss of entanglement, with the pulse further accentuating this dissipative trend. However, the persistence of high entropic uncertainty shows that even when NG and QD vanish, the system retains a residual quantum signature linked to memory effects and the intrinsic structure of quantum coherence.

	\section{Conclusion}
	
	In this work, we investigated the dynamics of quantum correlations in a bipartite two-qubit system by monitoring Negativity (NG), Quantum Discord (QD), and Quantum-Memory-Assisted Entropic Uncertainty (QM-EUR) under the influence of external pulses and various types of decoherence, including amplitude damping ($\gamma_{\rm amp}$), pure dephasing ($\gamma_{\rm deph}$), and pulse-induced dephasing ($G$). The initial states considered were maximally entangled ($\ket{\psi^+}$) and separable ($\ket{00}$), while different regimes of inter-qubit coupling ($J_{zz}$, $J_{xx}$), qubit energy splitting ($\epsilon$), and pulse parameters ($A_{\rm pulse}$, $\beta_{\rm pulse}$) were explored. Our results show that the initial state strongly governs both the robustness and generation of correlations. Bell states maintain high correlations but gradually decay, while separable states can temporarily develop entanglement and quantum correlations under strong or rapid pulses. Inter-qubit coupling and energy splitting $\epsilon$ critically influence the dynamics. In the weak-coupling regime ($\epsilon \ll 1$), all correlation measures display pronounced oscillations, allowing even separable states to acquire significant transient correlations under the pulse. In contrast, in the strong-coupling regime ($\epsilon \gg 1$), the Bell state retains a clear advantage, with NG and QD remaining high, while separable states fail to generate new correlations, indicating that strong energy splitting protects existing entanglement but suppresses correlation generation from unentangled states. A clear hierarchy of robustness emerges. NG is the most sensitive, decaying quickly under decoherence or intense pulses, QD persists longer, revealing nonclassical correlations independent of entanglement, while QM-EUR reflects residual quantum memory and entropic uncertainty, showing that quantum signatures survive even when NG and QD are weak. Pulse amplitude and width ($A_{\rm pulse}$, $\beta_{\rm pulse}$) effectively control correlation generation and dissipation, while the intensity of pulse-induced dephasing ($G$) modulates sustained oscillations versus rapid decoherence.
	
	Overall, this study highlights the crucial role of the initial state, interqubit coupling, energy splitting $\epsilon$, and pulse characteristics in preserving, creating, and modulating quantum correlations. These findings provide valuable guidance for the precise control of quantum resources and quantum memory, contributing to the development of robust devices and protocols for quantum information. They are in agreement with recent works, for instance, the generation and preservation of entanglement via external control fields in two-qubit systems under Markovian noise~\cite{ghorbani2025quantum}, the study of correlation dynamics in two coupled qubits driven by time-dependent fields~\cite{ghiu2020quantum}, and the protection of quantum discord using bang-bang pulse sequences in non-Markovian environments~\cite{xu2012protecting}.
	
	Several promising research directions emerge from this work. It would be interesting to study thermal effects and realistic environments on correlation dynamics~\cite{ghorbani2025quantum}, extend the analysis to multipartite or qubit–qutrit systems~\cite{ghiu2020quantum}, and experimentally validate predictions via artificial reservoirs or photonic simulators \cite{ferreira2021collapse}. Integrating artificial intelligence techniques for adaptive pulse control and exploring their impact on quantum metrology also offer promising avenues.
	
	These results demonstrate that precise control of external pulses enables effective modulation and preservation of quantum correlations, exploiting the system's intrinsic memory effects to mitigate decoherence. This approach highlights the potential of controlled pulses in the development of robust quantum systems, such as gate-based quantum computers, distributed quantum networks~\cite{gyongyosi2020decoherence, gyongyosi2021scalable}, and the emerging quantum internet~\cite{gyongyosi2022adaptive,gyongyosi2022advances}.

	\bibliography{references2.bib}

\begin{thebibliography}{48}%
\makeatletter
\providecommand \@ifxundefined [1]{%
 \@ifx{#1\undefined}
}%
\providecommand \@ifnum [1]{%
 \ifnum #1\expandafter \@firstoftwo
 \else \expandafter \@secondoftwo
 \fi
}%
\providecommand \@ifx [1]{%
 \ifx #1\expandafter \@firstoftwo
 \else \expandafter \@secondoftwo
 \fi
}%
\providecommand \natexlab [1]{#1}%
\providecommand \enquote  [1]{``#1''}%
\providecommand \bibnamefont  [1]{#1}%
\providecommand \bibfnamefont [1]{#1}%
\providecommand \citenamefont [1]{#1}%
\providecommand \href@noop [0]{\@secondoftwo}%
\providecommand \href [0]{\begingroup \@sanitize@url \@href}%
\providecommand \@href[1]{\@@startlink{#1}\@@href}%
\providecommand \@@href[1]{\endgroup#1\@@endlink}%
\providecommand \@sanitize@url [0]{\catcode `\\12\catcode `\$12\catcode
  `\&12\catcode `\#12\catcode `\^12\catcode `\_12\catcode `\%12\relax}%
\providecommand \@@startlink[1]{}%
\providecommand \@@endlink[0]{}%
\providecommand \url  [0]{\begingroup\@sanitize@url \@url }%
\providecommand \@url [1]{\endgroup\@href {#1}{\urlprefix }}%
\providecommand \urlprefix  [0]{URL }%
\providecommand \Eprint [0]{\href }%
\providecommand \doibase [0]{https://doi.org/}%
\providecommand \selectlanguage [0]{\@gobble}%
\providecommand \bibinfo  [0]{\@secondoftwo}%
\providecommand \bibfield  [0]{\@secondoftwo}%
\providecommand \translation [1]{[#1]}%
\providecommand \BibitemOpen [0]{}%
\providecommand \bibitemStop [0]{}%
\providecommand \bibitemNoStop [0]{.\EOS\space}%
\providecommand \EOS [0]{\spacefactor3000\relax}%
\providecommand \BibitemShut  [1]{\csname bibitem#1\endcsname}%
\let\auto@bib@innerbib\@empty
\bibitem [{\citenamefont {Nielsen}\ and\ \citenamefont
  {Chuang}(2010)}]{ref:nielsen_chuang}%
  \BibitemOpen
  \bibfield  {author} {\bibinfo {author} {\bibfnamefont {M.~A.}\ \bibnamefont
  {Nielsen}}\ and\ \bibinfo {author} {\bibfnamefont {I.~L.}\ \bibnamefont
  {Chuang}},\ }\href@noop {} {\emph {\bibinfo {title} {Quantum Computation and
  Quantum Information}}}\ (\bibinfo  {publisher} {Cambridge University Press},\
  \bibinfo {year} {2010})\BibitemShut {NoStop}%
\bibitem [{\citenamefont {Breuer}\ and\ \citenamefont
  {Petruccione}(2002)}]{breuer2002theory}%
  \BibitemOpen
  \bibfield  {author} {\bibinfo {author} {\bibfnamefont {H.-P.}\ \bibnamefont
  {Breuer}}\ and\ \bibinfo {author} {\bibfnamefont {F.}~\bibnamefont
  {Petruccione}},\ }\href@noop {} {\emph {\bibinfo {title} {The theory of open
  quantum systems}}}\ (\bibinfo  {publisher} {OUP Oxford},\ \bibinfo {year}
  {2002})\BibitemShut {NoStop}%
\bibitem [{\citenamefont {Terhal}(2015)}]{terhal2015quantum}%
  \BibitemOpen
  \bibfield  {author} {\bibinfo {author} {\bibfnamefont {B.~M.}\ \bibnamefont
  {Terhal}},\ }\bibfield  {title} {\bibinfo {title} {Quantum error correction
  for quantum memories},\ }\href@noop {} {\bibfield  {journal} {\bibinfo
  {journal} {Reviews of Modern Physics}\ }\textbf {\bibinfo {volume} {87}},\
  \bibinfo {pages} {307} (\bibinfo {year} {2015})}\BibitemShut {NoStop}%
\bibitem [{\citenamefont {Viola}\ and\ \citenamefont
  {Lloyd}(1998)}]{viola1998dynamical}%
  \BibitemOpen
  \bibfield  {author} {\bibinfo {author} {\bibfnamefont {L.}~\bibnamefont
  {Viola}}\ and\ \bibinfo {author} {\bibfnamefont {S.}~\bibnamefont {Lloyd}},\
  }\bibfield  {title} {\bibinfo {title} {Dynamical suppression of decoherence
  in two-state quantum systems},\ }\href@noop {} {\bibfield  {journal}
  {\bibinfo  {journal} {Physical Review A}\ }\textbf {\bibinfo {volume} {58}},\
  \bibinfo {pages} {2733} (\bibinfo {year} {1998})}\BibitemShut {NoStop}%
\bibitem [{\citenamefont {Temme}\ \emph {et~al.}(2017)\citenamefont {Temme},
  \citenamefont {Bravyi},\ and\ \citenamefont {Gambetta}}]{temme2017error}%
  \BibitemOpen
  \bibfield  {author} {\bibinfo {author} {\bibfnamefont {K.}~\bibnamefont
  {Temme}}, \bibinfo {author} {\bibfnamefont {S.}~\bibnamefont {Bravyi}},\ and\
  \bibinfo {author} {\bibfnamefont {J.~M.}\ \bibnamefont {Gambetta}},\
  }\bibfield  {title} {\bibinfo {title} {Error mitigation for short-depth
  quantum circuits},\ }\href@noop {} {\bibfield  {journal} {\bibinfo  {journal}
  {Physical review letters}\ }\textbf {\bibinfo {volume} {119}},\ \bibinfo
  {pages} {180509} (\bibinfo {year} {2017})}\BibitemShut {NoStop}%
\bibitem [{\citenamefont {Glaser}\ \emph {et~al.}(2015)\citenamefont {Glaser},
  \citenamefont {Boscain}, \citenamefont {Calarco}, \citenamefont {Koch},
  \citenamefont {K{\"o}ckenberger}, \citenamefont {Kosloff}, \citenamefont
  {Kuprov}, \citenamefont {Luy}, \citenamefont {Schirmer}, \citenamefont
  {Schulte-Herbr{\"u}ggen} \emph {et~al.}}]{glaser2015training}%
  \BibitemOpen
  \bibfield  {author} {\bibinfo {author} {\bibfnamefont {S.~J.}\ \bibnamefont
  {Glaser}}, \bibinfo {author} {\bibfnamefont {U.}~\bibnamefont {Boscain}},
  \bibinfo {author} {\bibfnamefont {T.}~\bibnamefont {Calarco}}, \bibinfo
  {author} {\bibfnamefont {C.~P.}\ \bibnamefont {Koch}}, \bibinfo {author}
  {\bibfnamefont {W.}~\bibnamefont {K{\"o}ckenberger}}, \bibinfo {author}
  {\bibfnamefont {R.}~\bibnamefont {Kosloff}}, \bibinfo {author} {\bibfnamefont
  {I.}~\bibnamefont {Kuprov}}, \bibinfo {author} {\bibfnamefont
  {B.}~\bibnamefont {Luy}}, \bibinfo {author} {\bibfnamefont {S.}~\bibnamefont
  {Schirmer}}, \bibinfo {author} {\bibfnamefont {T.}~\bibnamefont
  {Schulte-Herbr{\"u}ggen}}, \emph {et~al.},\ }\bibfield  {title} {\bibinfo
  {title} {Training schr{\"o}dinger’s cat: Quantum optimal control: Strategic
  report on current status, visions and goals for research in europe},\
  }\href@noop {} {\bibfield  {journal} {\bibinfo  {journal} {The European
  Physical Journal D}\ }\textbf {\bibinfo {volume} {69}},\ \bibinfo {pages}
  {279} (\bibinfo {year} {2015})}\BibitemShut {NoStop}%
\bibitem [{\citenamefont {Amico}\ \emph {et~al.}(2008)\citenamefont {Amico},
  \citenamefont {Fazio}, \citenamefont {Osterloh},\ and\ \citenamefont
  {Vedral}}]{amico2008entanglement}%
  \BibitemOpen
  \bibfield  {author} {\bibinfo {author} {\bibfnamefont {L.}~\bibnamefont
  {Amico}}, \bibinfo {author} {\bibfnamefont {R.}~\bibnamefont {Fazio}},
  \bibinfo {author} {\bibfnamefont {A.}~\bibnamefont {Osterloh}},\ and\
  \bibinfo {author} {\bibfnamefont {V.}~\bibnamefont {Vedral}},\ }\bibfield
  {title} {\bibinfo {title} {Entanglement in many-body systems},\ }\href@noop
  {} {\bibfield  {journal} {\bibinfo  {journal} {Reviews of modern physics}\
  }\textbf {\bibinfo {volume} {80}},\ \bibinfo {pages} {517} (\bibinfo {year}
  {2008})}\BibitemShut {NoStop}%
\bibitem [{\citenamefont {Blais}\ \emph {et~al.}(2021)\citenamefont {Blais},
  \citenamefont {Grimsmo}, \citenamefont {Girvin},\ and\ \citenamefont
  {Wallraff}}]{blais2021circuit}%
  \BibitemOpen
  \bibfield  {author} {\bibinfo {author} {\bibfnamefont {A.}~\bibnamefont
  {Blais}}, \bibinfo {author} {\bibfnamefont {A.~L.}\ \bibnamefont {Grimsmo}},
  \bibinfo {author} {\bibfnamefont {S.~M.}\ \bibnamefont {Girvin}},\ and\
  \bibinfo {author} {\bibfnamefont {A.}~\bibnamefont {Wallraff}},\ }\bibfield
  {title} {\bibinfo {title} {Circuit quantum electrodynamics},\ }\href@noop {}
  {\bibfield  {journal} {\bibinfo  {journal} {Reviews of Modern Physics}\
  }\textbf {\bibinfo {volume} {93}},\ \bibinfo {pages} {025005} (\bibinfo
  {year} {2021})}\BibitemShut {NoStop}%
\bibitem [{\citenamefont {Vinjanampathy}\ and\ \citenamefont
  {Anders}(2016)}]{vinjanampathy2016quantum}%
  \BibitemOpen
  \bibfield  {author} {\bibinfo {author} {\bibfnamefont {S.}~\bibnamefont
  {Vinjanampathy}}\ and\ \bibinfo {author} {\bibfnamefont {J.}~\bibnamefont
  {Anders}},\ }\bibfield  {title} {\bibinfo {title} {Quantum thermodynamics},\
  }\href@noop {} {\bibfield  {journal} {\bibinfo  {journal} {Contemporary
  Physics}\ }\textbf {\bibinfo {volume} {57}},\ \bibinfo {pages} {545}
  (\bibinfo {year} {2016})}\BibitemShut {NoStop}%
\bibitem [{\citenamefont {Nandkishore}\ and\ \citenamefont
  {Huse}(2015)}]{nandkishore2015many}%
  \BibitemOpen
  \bibfield  {author} {\bibinfo {author} {\bibfnamefont {R.}~\bibnamefont
  {Nandkishore}}\ and\ \bibinfo {author} {\bibfnamefont {D.~A.}\ \bibnamefont
  {Huse}},\ }\bibfield  {title} {\bibinfo {title} {Many-body localization and
  thermalization in quantum statistical mechanics},\ }\href@noop {} {\bibfield
  {journal} {\bibinfo  {journal} {Annu. Rev. Condens. Matter Phys.}\ }\textbf
  {\bibinfo {volume} {6}},\ \bibinfo {pages} {15} (\bibinfo {year}
  {2015})}\BibitemShut {NoStop}%
\bibitem [{\citenamefont {Martinis}\ and\ \citenamefont
  {Geller}(2014)}]{martinis2014fast}%
  \BibitemOpen
  \bibfield  {author} {\bibinfo {author} {\bibfnamefont {J.~M.}\ \bibnamefont
  {Martinis}}\ and\ \bibinfo {author} {\bibfnamefont {M.~R.}\ \bibnamefont
  {Geller}},\ }\bibfield  {title} {\bibinfo {title} {Fast adiabatic qubit gates
  using only $\sigma$ z control},\ }\href@noop {} {\bibfield  {journal}
  {\bibinfo  {journal} {Physical Review A}\ }\textbf {\bibinfo {volume} {90}},\
  \bibinfo {pages} {022307} (\bibinfo {year} {2014})}\BibitemShut {NoStop}%
\bibitem [{\citenamefont {Caneva}\ \emph {et~al.}(2011)\citenamefont {Caneva},
  \citenamefont {Calarco},\ and\ \citenamefont
  {Montangero}}]{caneva2011chopped}%
  \BibitemOpen
  \bibfield  {author} {\bibinfo {author} {\bibfnamefont {T.}~\bibnamefont
  {Caneva}}, \bibinfo {author} {\bibfnamefont {T.}~\bibnamefont {Calarco}},\
  and\ \bibinfo {author} {\bibfnamefont {S.}~\bibnamefont {Montangero}},\
  }\bibfield  {title} {\bibinfo {title} {Chopped random-basis quantum
  optimization},\ }\href@noop {} {\bibfield  {journal} {\bibinfo  {journal}
  {Physical Review A—Atomic, Molecular, and Optical Physics}\ }\textbf
  {\bibinfo {volume} {84}},\ \bibinfo {pages} {022326} (\bibinfo {year}
  {2011})}\BibitemShut {NoStop}%
\bibitem [{\citenamefont {Bialczak}\ \emph {et~al.}(2011)\citenamefont
  {Bialczak}, \citenamefont {Ansmann}, \citenamefont {Hofheinz}, \citenamefont
  {Lenander}, \citenamefont {Lucero}, \citenamefont {Neeley}, \citenamefont
  {O’Connell}, \citenamefont {Sank}, \citenamefont {Wang}, \citenamefont
  {Weides} \emph {et~al.}}]{bialczak2011fast}%
  \BibitemOpen
  \bibfield  {author} {\bibinfo {author} {\bibfnamefont {R.}~\bibnamefont
  {Bialczak}}, \bibinfo {author} {\bibfnamefont {M.}~\bibnamefont {Ansmann}},
  \bibinfo {author} {\bibfnamefont {M.}~\bibnamefont {Hofheinz}}, \bibinfo
  {author} {\bibfnamefont {M.}~\bibnamefont {Lenander}}, \bibinfo {author}
  {\bibfnamefont {E.}~\bibnamefont {Lucero}}, \bibinfo {author} {\bibfnamefont
  {M.}~\bibnamefont {Neeley}}, \bibinfo {author} {\bibfnamefont
  {A.}~\bibnamefont {O’Connell}}, \bibinfo {author} {\bibfnamefont
  {D.}~\bibnamefont {Sank}}, \bibinfo {author} {\bibfnamefont {H.}~\bibnamefont
  {Wang}}, \bibinfo {author} {\bibfnamefont {M.}~\bibnamefont {Weides}}, \emph
  {et~al.},\ }\bibfield  {title} {\bibinfo {title} {Fast tunable coupler for
  superconducting qubits},\ }\href@noop {} {\bibfield  {journal} {\bibinfo
  {journal} {Physical review letters}\ }\textbf {\bibinfo {volume} {106}},\
  \bibinfo {pages} {060501} (\bibinfo {year} {2011})}\BibitemShut {NoStop}%
\bibitem [{\citenamefont {Eckardt}(2017)}]{eckardt2017colloquium}%
  \BibitemOpen
  \bibfield  {author} {\bibinfo {author} {\bibfnamefont {A.}~\bibnamefont
  {Eckardt}},\ }\bibfield  {title} {\bibinfo {title} {Colloquium: Atomic
  quantum gases in periodically driven optical lattices},\ }\href@noop {}
  {\bibfield  {journal} {\bibinfo  {journal} {Reviews of Modern Physics}\
  }\textbf {\bibinfo {volume} {89}},\ \bibinfo {pages} {011004} (\bibinfo
  {year} {2017})}\BibitemShut {NoStop}%
\bibitem [{\citenamefont {Zurek}(2003)}]{zurek2003decoherence}%
  \BibitemOpen
  \bibfield  {author} {\bibinfo {author} {\bibfnamefont {W.~H.}\ \bibnamefont
  {Zurek}},\ }\bibfield  {title} {\bibinfo {title} {Decoherence, einselection,
  and the quantum origins of the classical},\ }\href@noop {} {\bibfield
  {journal} {\bibinfo  {journal} {Reviews of modern physics}\ }\textbf
  {\bibinfo {volume} {75}},\ \bibinfo {pages} {715} (\bibinfo {year}
  {2003})}\BibitemShut {NoStop}%
\bibitem [{\citenamefont {Lidar}\ and\ \citenamefont
  {Brun}(2013)}]{lidar2013quantum}%
  \BibitemOpen
  \bibfield  {author} {\bibinfo {author} {\bibfnamefont {D.~A.}\ \bibnamefont
  {Lidar}}\ and\ \bibinfo {author} {\bibfnamefont {T.~A.}\ \bibnamefont
  {Brun}},\ }\href@noop {} {\emph {\bibinfo {title} {Quantum error
  correction}}}\ (\bibinfo  {publisher} {Cambridge university press},\ \bibinfo
  {year} {2013})\BibitemShut {NoStop}%
\bibitem [{\citenamefont {Lidar}\ \emph {et~al.}(1998)\citenamefont {Lidar},
  \citenamefont {Chuang},\ and\ \citenamefont {Whaley}}]{lidar1998decoherence}%
  \BibitemOpen
  \bibfield  {author} {\bibinfo {author} {\bibfnamefont {D.~A.}\ \bibnamefont
  {Lidar}}, \bibinfo {author} {\bibfnamefont {I.~L.}\ \bibnamefont {Chuang}},\
  and\ \bibinfo {author} {\bibfnamefont {K.~B.}\ \bibnamefont {Whaley}},\
  }\bibfield  {title} {\bibinfo {title} {Decoherence-free subspaces for quantum
  computation},\ }\href@noop {} {\bibfield  {journal} {\bibinfo  {journal}
  {Physical Review Letters}\ }\textbf {\bibinfo {volume} {81}},\ \bibinfo
  {pages} {2594} (\bibinfo {year} {1998})}\BibitemShut {NoStop}%
\bibitem [{\citenamefont {Schlosshauer}(2007)}]{schlosshauer2007decoherence}%
  \BibitemOpen
  \bibfield  {author} {\bibinfo {author} {\bibfnamefont {M.}~\bibnamefont
  {Schlosshauer}},\ }\href@noop {} {\emph {\bibinfo {title} {Decoherence and
  the quantum-to-classical transition}}}\ (\bibinfo  {publisher} {Springer},\
  \bibinfo {year} {2007})\BibitemShut {NoStop}%
\bibitem [{\citenamefont {Yu}\ and\ \citenamefont
  {Eberly}(2009)}]{yu2009sudden}%
  \BibitemOpen
  \bibfield  {author} {\bibinfo {author} {\bibfnamefont {T.}~\bibnamefont
  {Yu}}\ and\ \bibinfo {author} {\bibfnamefont {J.}~\bibnamefont {Eberly}},\
  }\bibfield  {title} {\bibinfo {title} {Sudden death of entanglement},\
  }\href@noop {} {\bibfield  {journal} {\bibinfo  {journal} {Science}\ }\textbf
  {\bibinfo {volume} {323}},\ \bibinfo {pages} {598} (\bibinfo {year}
  {2009})}\BibitemShut {NoStop}%
\bibitem [{\citenamefont {Vidal}\ and\ \citenamefont
  {Werner}(2002{\natexlab{a}})}]{vidal2002computable}%
  \BibitemOpen
  \bibfield  {author} {\bibinfo {author} {\bibfnamefont {G.}~\bibnamefont
  {Vidal}}\ and\ \bibinfo {author} {\bibfnamefont {R.~F.}\ \bibnamefont
  {Werner}},\ }\bibfield  {title} {\bibinfo {title} {Computable measure of
  entanglement},\ }\href@noop {} {\bibfield  {journal} {\bibinfo  {journal}
  {Physical Review A}\ }\textbf {\bibinfo {volume} {65}},\ \bibinfo {pages}
  {032314} (\bibinfo {year} {2002}{\natexlab{a}})}\BibitemShut {NoStop}%
\bibitem [{\citenamefont {Ollivier}\ and\ \citenamefont
  {Zurek}(2001{\natexlab{a}})}]{ollivier2001quantum}%
  \BibitemOpen
  \bibfield  {author} {\bibinfo {author} {\bibfnamefont {H.}~\bibnamefont
  {Ollivier}}\ and\ \bibinfo {author} {\bibfnamefont {W.~H.}\ \bibnamefont
  {Zurek}},\ }\bibfield  {title} {\bibinfo {title} {Quantum discord: a measure
  of the quantumness of correlations},\ }\href@noop {} {\bibfield  {journal}
  {\bibinfo  {journal} {Physical review letters}\ }\textbf {\bibinfo {volume}
  {88}},\ \bibinfo {pages} {017901} (\bibinfo {year}
  {2001}{\natexlab{a}})}\BibitemShut {NoStop}%
\bibitem [{\citenamefont {Modi}\ \emph {et~al.}(2012)\citenamefont {Modi},
  \citenamefont {Brodutch}, \citenamefont {Cable}, \citenamefont {Paterek},\
  and\ \citenamefont {Vedral}}]{modi2012classical}%
  \BibitemOpen
  \bibfield  {author} {\bibinfo {author} {\bibfnamefont {K.}~\bibnamefont
  {Modi}}, \bibinfo {author} {\bibfnamefont {A.}~\bibnamefont {Brodutch}},
  \bibinfo {author} {\bibfnamefont {H.}~\bibnamefont {Cable}}, \bibinfo
  {author} {\bibfnamefont {T.}~\bibnamefont {Paterek}},\ and\ \bibinfo {author}
  {\bibfnamefont {V.}~\bibnamefont {Vedral}},\ }\bibfield  {title} {\bibinfo
  {title} {The classical-quantum boundary for correlations:<? format?> discord
  and related measures},\ }\href@noop {} {\bibfield  {journal} {\bibinfo
  {journal} {Reviews of Modern Physics}\ }\textbf {\bibinfo {volume} {84}},\
  \bibinfo {pages} {1655} (\bibinfo {year} {2012})}\BibitemShut {NoStop}%
\bibitem [{\citenamefont {Breuer}\ \emph {et~al.}(2016)\citenamefont {Breuer},
  \citenamefont {Laine}, \citenamefont {Piilo},\ and\ \citenamefont
  {Vacchini}}]{breuer2016colloquium}%
  \BibitemOpen
  \bibfield  {author} {\bibinfo {author} {\bibfnamefont {H.-P.}\ \bibnamefont
  {Breuer}}, \bibinfo {author} {\bibfnamefont {E.-M.}\ \bibnamefont {Laine}},
  \bibinfo {author} {\bibfnamefont {J.}~\bibnamefont {Piilo}},\ and\ \bibinfo
  {author} {\bibfnamefont {B.}~\bibnamefont {Vacchini}},\ }\bibfield  {title}
  {\bibinfo {title} {Colloquium: Non-markovian dynamics in open quantum
  systems},\ }\href@noop {} {\bibfield  {journal} {\bibinfo  {journal} {Reviews
  of Modern Physics}\ }\textbf {\bibinfo {volume} {88}},\ \bibinfo {pages}
  {021002} (\bibinfo {year} {2016})}\BibitemShut {NoStop}%
\bibitem [{\citenamefont {Datta}\ \emph {et~al.}(2008)\citenamefont {Datta},
  \citenamefont {Shaji},\ and\ \citenamefont {Caves}}]{datta2008quantum}%
  \BibitemOpen
  \bibfield  {author} {\bibinfo {author} {\bibfnamefont {A.}~\bibnamefont
  {Datta}}, \bibinfo {author} {\bibfnamefont {A.}~\bibnamefont {Shaji}},\ and\
  \bibinfo {author} {\bibfnamefont {C.~M.}\ \bibnamefont {Caves}},\ }\bibfield
  {title} {\bibinfo {title} {Quantum discord and the power of one qubit},\
  }\href@noop {} {\bibfield  {journal} {\bibinfo  {journal} {Physical review
  letters}\ }\textbf {\bibinfo {volume} {100}},\ \bibinfo {pages} {050502}
  (\bibinfo {year} {2008})}\BibitemShut {NoStop}%
\bibitem [{\citenamefont {De~Vega}\ and\ \citenamefont
  {Alonso}(2017)}]{de2017dynamics}%
  \BibitemOpen
  \bibfield  {author} {\bibinfo {author} {\bibfnamefont {I.}~\bibnamefont
  {De~Vega}}\ and\ \bibinfo {author} {\bibfnamefont {D.}~\bibnamefont
  {Alonso}},\ }\bibfield  {title} {\bibinfo {title} {Dynamics of non-markovian
  open quantum systems},\ }\href@noop {} {\bibfield  {journal} {\bibinfo
  {journal} {Reviews of Modern Physics}\ }\textbf {\bibinfo {volume} {89}},\
  \bibinfo {pages} {015001} (\bibinfo {year} {2017})}\BibitemShut {NoStop}%
\bibitem [{\citenamefont {de~Clercq}\ \emph {et~al.}(2016)\citenamefont
  {de~Clercq}, \citenamefont {Oswald}, \citenamefont {Flühmann}, \citenamefont
  {Keitch}, \citenamefont {Kienzler}, \citenamefont {Lo}, \citenamefont
  {Marinelli}, \citenamefont {Nadlinger}, \citenamefont {Negnevitsky},\ and\
  \citenamefont {Home}}]{deClercq2016}%
  \BibitemOpen
  \bibfield  {author} {\bibinfo {author} {\bibfnamefont {L.~E.}\ \bibnamefont
  {de~Clercq}}, \bibinfo {author} {\bibfnamefont {R.}~\bibnamefont {Oswald}},
  \bibinfo {author} {\bibfnamefont {C.}~\bibnamefont {Flühmann}}, \bibinfo
  {author} {\bibfnamefont {B.}~\bibnamefont {Keitch}}, \bibinfo {author}
  {\bibfnamefont {D.}~\bibnamefont {Kienzler}}, \bibinfo {author}
  {\bibfnamefont {H.-Y.}\ \bibnamefont {Lo}}, \bibinfo {author} {\bibfnamefont
  {M.}~\bibnamefont {Marinelli}}, \bibinfo {author} {\bibfnamefont
  {D.}~\bibnamefont {Nadlinger}}, \bibinfo {author} {\bibfnamefont
  {V.}~\bibnamefont {Negnevitsky}},\ and\ \bibinfo {author} {\bibfnamefont
  {J.~P.}\ \bibnamefont {Home}},\ }\bibfield  {title} {\bibinfo {title}
  {Estimation of a general time-dependent hamiltonian for a single qubit},\
  }\href {https://doi.org/10.1038/ncomms11218} {\bibfield  {journal} {\bibinfo
  {journal} {Nature Communications}\ }\textbf {\bibinfo {volume} {7}},\
  \bibinfo {pages} {11218} (\bibinfo {year} {2016})},\ \bibinfo {note}
  {published 14 April 2016}\BibitemShut {NoStop}%
\bibitem [{\citenamefont {Vaicaitis}(nd)}]{Vaicaitis_Rimas_Pauli_operators}%
  \BibitemOpen
  \bibfield  {author} {\bibinfo {author} {\bibfnamefont {R.}~\bibnamefont
  {Vaicaitis}},\ }\href@noop {} {\bibinfo {title} {Pauli operators in quantum
  computing}},\ \bibinfo {howpublished} {Online lecture notes, Columbia
  University} (\bibinfo {year} {n.d.}),\ \bibinfo {note} {pDF available at
  \texttt{https://www.math.columbia.edu/~plei/docs/quantum/rimas.pdf}}\BibitemShut
  {NoStop}%
\bibitem [{\citenamefont {Majeed}\ and\ \citenamefont
  {Chaudhry}(2020)}]{majeed2020quantumzenoantizenoeffects}%
  \BibitemOpen
  \bibfield  {author} {\bibinfo {author} {\bibfnamefont {M.}~\bibnamefont
  {Majeed}}\ and\ \bibinfo {author} {\bibfnamefont {A.~Z.}\ \bibnamefont
  {Chaudhry}},\ }\href {https://arxiv.org/abs/2012.15040} {\bibinfo {title}
  {The quantum zeno and anti-zeno effects with driving fields in the weak and
  strong coupling regimes}} (\bibinfo {year} {2020}),\ \Eprint
  {https://arxiv.org/abs/2012.15040} {arXiv:2012.15040 [quant-ph]} \BibitemShut
  {NoStop}%
\bibitem [{\citenamefont {Kofman}\ and\ \citenamefont
  {Kurizki}(2001)}]{Kofman_2001}%
  \BibitemOpen
  \bibfield  {author} {\bibinfo {author} {\bibfnamefont {A.~G.}\ \bibnamefont
  {Kofman}}\ and\ \bibinfo {author} {\bibfnamefont {G.}~\bibnamefont
  {Kurizki}},\ }\bibfield  {title} {\bibinfo {title} {Universal dynamical
  control of quantum mechanical decay: Modulation of the coupling to the
  continuum},\ }\bibfield  {journal} {\bibinfo  {journal} {Physical Review
  Letters}\ }\textbf {\bibinfo {volume} {87}},\ \href
  {https://doi.org/10.1103/physrevlett.87.270405}
  {10.1103/physrevlett.87.270405} (\bibinfo {year} {2001})\BibitemShut
  {NoStop}%
\bibitem [{\citenamefont {Gordon}\ \emph {et~al.}(2008)\citenamefont {Gordon},
  \citenamefont {Kurizki},\ and\ \citenamefont {Lidar}}]{Gordon_2008}%
  \BibitemOpen
  \bibfield  {author} {\bibinfo {author} {\bibfnamefont {G.}~\bibnamefont
  {Gordon}}, \bibinfo {author} {\bibfnamefont {G.}~\bibnamefont {Kurizki}},\
  and\ \bibinfo {author} {\bibfnamefont {D.~A.}\ \bibnamefont {Lidar}},\
  }\bibfield  {title} {\bibinfo {title} {Optimal dynamical decoherence control
  of a qubit},\ }\bibfield  {journal} {\bibinfo  {journal} {Physical Review
  Letters}\ }\textbf {\bibinfo {volume} {101}},\ \href
  {https://doi.org/10.1103/physrevlett.101.010403}
  {10.1103/physrevlett.101.010403} (\bibinfo {year} {2008})\BibitemShut
  {NoStop}%
\bibitem [{\citenamefont {Gordon}\ and\ \citenamefont
  {Kurizki}(2006)}]{Gordon2006Universal}%
  \BibitemOpen
  \bibfield  {author} {\bibinfo {author} {\bibfnamefont {G.}~\bibnamefont
  {Gordon}}\ and\ \bibinfo {author} {\bibfnamefont {G.}~\bibnamefont
  {Kurizki}},\ }\bibfield  {title} {\bibinfo {title} {Universal dephasing
  control during quantum computation},\ }\href
  {https://doi.org/10.1103/PhysRevLett.97.110503} {\bibfield  {journal}
  {\bibinfo  {journal} {Physical Review Letters}\ }\textbf {\bibinfo {volume}
  {97}},\ \bibinfo {pages} {110503} (\bibinfo {year} {2006})}\BibitemShut
  {NoStop}%
\bibitem [{\citenamefont {Vitanov}\ \emph {et~al.}(2017)\citenamefont
  {Vitanov}, \citenamefont {Rangelov}, \citenamefont {Shore},\ and\
  \citenamefont {Bergmann}}]{vitanov2017stimulated}%
  \BibitemOpen
  \bibfield  {author} {\bibinfo {author} {\bibfnamefont {N.~V.}\ \bibnamefont
  {Vitanov}}, \bibinfo {author} {\bibfnamefont {A.~A.}\ \bibnamefont
  {Rangelov}}, \bibinfo {author} {\bibfnamefont {B.~W.}\ \bibnamefont
  {Shore}},\ and\ \bibinfo {author} {\bibfnamefont {K.}~\bibnamefont
  {Bergmann}},\ }\bibfield  {title} {\bibinfo {title} {Stimulated raman
  adiabatic passage in physics, chemistry, and beyond},\ }\href
  {https://doi.org/10.1103/RevModPhys.89.015006} {\bibfield  {journal}
  {\bibinfo  {journal} {Reviews of Modern Physics}\ }\textbf {\bibinfo {volume}
  {89}},\ \bibinfo {pages} {015006} (\bibinfo {year} {2017})}\BibitemShut
  {NoStop}%
\bibitem [{\citenamefont {Motzoi}\ \emph {et~al.}(2009)\citenamefont {Motzoi},
  \citenamefont {Gambetta}, \citenamefont {Rebentrost},\ and\ \citenamefont
  {Wilhelm}}]{motzoi2009simple}%
  \BibitemOpen
  \bibfield  {author} {\bibinfo {author} {\bibfnamefont {F.}~\bibnamefont
  {Motzoi}}, \bibinfo {author} {\bibfnamefont {J.~M.}\ \bibnamefont
  {Gambetta}}, \bibinfo {author} {\bibfnamefont {P.}~\bibnamefont
  {Rebentrost}},\ and\ \bibinfo {author} {\bibfnamefont {F.~K.}\ \bibnamefont
  {Wilhelm}},\ }\bibfield  {title} {\bibinfo {title} {Simple pulses for
  elimination of leakage in weakly nonlinear qubits},\ }\href@noop {}
  {\bibfield  {journal} {\bibinfo  {journal} {Physical Review Letters}\
  }\textbf {\bibinfo {volume} {103}},\ \bibinfo {pages} {110501} (\bibinfo
  {year} {2009})}\BibitemShut {NoStop}%
\bibitem [{\citenamefont {Krantz}\ \emph {et~al.}(2019)\citenamefont {Krantz},
  \citenamefont {Kjaergaard}, \citenamefont {Yan}, \citenamefont {Orlando},
  \citenamefont {Gustavsson},\ and\ \citenamefont
  {Oliver}}]{krantz2019quantum}%
  \BibitemOpen
  \bibfield  {author} {\bibinfo {author} {\bibfnamefont {P.}~\bibnamefont
  {Krantz}}, \bibinfo {author} {\bibfnamefont {M.}~\bibnamefont {Kjaergaard}},
  \bibinfo {author} {\bibfnamefont {F.}~\bibnamefont {Yan}}, \bibinfo {author}
  {\bibfnamefont {T.~P.}\ \bibnamefont {Orlando}}, \bibinfo {author}
  {\bibfnamefont {S.}~\bibnamefont {Gustavsson}},\ and\ \bibinfo {author}
  {\bibfnamefont {W.~D.}\ \bibnamefont {Oliver}},\ }\bibfield  {title}
  {\bibinfo {title} {A quantum engineer’s guide to superconducting qubits},\
  }\href@noop {} {\bibfield  {journal} {\bibinfo  {journal} {Applied Physics
  Reviews}\ }\textbf {\bibinfo {volume} {6}},\ \bibinfo {pages} {021318}
  (\bibinfo {year} {2019})}\BibitemShut {NoStop}%
\bibitem [{\citenamefont {Clarke}\ and\ \citenamefont
  {Wilhelm}(2008)}]{clarke2008superconducting}%
  \BibitemOpen
  \bibfield  {author} {\bibinfo {author} {\bibfnamefont {J.}~\bibnamefont
  {Clarke}}\ and\ \bibinfo {author} {\bibfnamefont {F.~K.}\ \bibnamefont
  {Wilhelm}},\ }\bibfield  {title} {\bibinfo {title} {Superconducting qubits},\
  }\href@noop {} {\bibfield  {journal} {\bibinfo  {journal} {Nature}\ }\textbf
  {\bibinfo {volume} {453}},\ \bibinfo {pages} {1031} (\bibinfo {year}
  {2008})}\BibitemShut {NoStop}%
\bibitem [{\citenamefont {Bell}(1964)}]{Bell_1964}%
  \BibitemOpen
  \bibfield  {author} {\bibinfo {author} {\bibfnamefont {J.~S.}\ \bibnamefont
  {Bell}},\ }\bibfield  {title} {\bibinfo {title} {On the einstein podolsky
  rosen paradox},\ }\href@noop {} {\bibfield  {journal} {\bibinfo  {journal}
  {Physics}\ }\textbf {\bibinfo {volume} {1}},\ \bibinfo {pages} {195}
  (\bibinfo {year} {1964})}\BibitemShut {NoStop}%
\bibitem [{\citenamefont {Vidal}\ and\ \citenamefont
  {Werner}(2002{\natexlab{b}})}]{Vidal_2002}%
  \BibitemOpen
  \bibfield  {author} {\bibinfo {author} {\bibfnamefont {G.}~\bibnamefont
  {Vidal}}\ and\ \bibinfo {author} {\bibfnamefont {R.~F.}\ \bibnamefont
  {Werner}},\ }\bibfield  {title} {\bibinfo {title} {A classical separability
  criterion for continuous variable quantum states},\ }\href@noop {} {\bibfield
   {journal} {\bibinfo  {journal} {Phys. Rev. A}\ }\textbf {\bibinfo {volume}
  {65}},\ \bibinfo {pages} {032314} (\bibinfo {year}
  {2002}{\natexlab{b}})}\BibitemShut {NoStop}%
\bibitem [{\citenamefont {Ollivier}\ and\ \citenamefont
  {Zurek}(2001{\natexlab{b}})}]{Ollivier_2001}%
  \BibitemOpen
  \bibfield  {author} {\bibinfo {author} {\bibfnamefont {H.}~\bibnamefont
  {Ollivier}}\ and\ \bibinfo {author} {\bibfnamefont {W.~H.}\ \bibnamefont
  {Zurek}},\ }\bibfield  {title} {\bibinfo {title} {Quantum discord: A measure
  of the quantumness of correlations},\ }\href@noop {} {\bibfield  {journal}
  {\bibinfo  {journal} {Phys. Rev. Lett.}\ }\textbf {\bibinfo {volume} {88}},\
  \bibinfo {pages} {017901} (\bibinfo {year} {2001}{\natexlab{b}})}\BibitemShut
  {NoStop}%
\bibitem [{\citenamefont {Berta}\ \emph {et~al.}(2010)\citenamefont {Berta},
  \citenamefont {Christandl}, \citenamefont {Colbeck}, \citenamefont {Renes},\
  and\ \citenamefont {Renner}}]{Berta2010}%
  \BibitemOpen
  \bibfield  {author} {\bibinfo {author} {\bibfnamefont {M.}~\bibnamefont
  {Berta}}, \bibinfo {author} {\bibfnamefont {M.}~\bibnamefont {Christandl}},
  \bibinfo {author} {\bibfnamefont {R.}~\bibnamefont {Colbeck}}, \bibinfo
  {author} {\bibfnamefont {J.~M.}\ \bibnamefont {Renes}},\ and\ \bibinfo
  {author} {\bibfnamefont {R.}~\bibnamefont {Renner}},\ }\bibfield  {title}
  {\bibinfo {title} {The uncertainty principle in the presence of quantum
  memory},\ }\href {https://doi.org/10.1038/nphys1734} {\bibfield  {journal}
  {\bibinfo  {journal} {Nature Physics}\ }\textbf {\bibinfo {volume} {6}},\
  \bibinfo {pages} {659} (\bibinfo {year} {2010})}\BibitemShut {NoStop}%
\bibitem [{\citenamefont {Coles}\ \emph {et~al.}(2017)\citenamefont {Coles},
  \citenamefont {Berta}, \citenamefont {Tomamichel},\ and\ \citenamefont
  {Wehner}}]{Coles2017}%
  \BibitemOpen
  \bibfield  {author} {\bibinfo {author} {\bibfnamefont {P.~J.}\ \bibnamefont
  {Coles}}, \bibinfo {author} {\bibfnamefont {M.}~\bibnamefont {Berta}},
  \bibinfo {author} {\bibfnamefont {M.}~\bibnamefont {Tomamichel}},\ and\
  \bibinfo {author} {\bibfnamefont {S.}~\bibnamefont {Wehner}},\ }\bibfield
  {title} {\bibinfo {title} {Entropic uncertainty relations and their
  applications},\ }\href {https://doi.org/10.1103/RevModPhys.89.015002}
  {\bibfield  {journal} {\bibinfo  {journal} {Reviews of Modern Physics}\
  }\textbf {\bibinfo {volume} {89}},\ \bibinfo {pages} {015002} (\bibinfo
  {year} {2017})}\BibitemShut {NoStop}%
\bibitem [{\citenamefont {Ghorbani}\ and\ \citenamefont
  {Rezaee}(2025)}]{ghorbani2025quantum}%
  \BibitemOpen
  \bibfield  {author} {\bibinfo {author} {\bibfnamefont {K.}~\bibnamefont
  {Ghorbani}}\ and\ \bibinfo {author} {\bibfnamefont {R.}~\bibnamefont
  {Rezaee}},\ }\bibfield  {title} {\bibinfo {title} {Quantum entanglement
  dynamics and concurrence preservation in a noisy two-qubit system with
  external control field},\ }\href@noop {} {\bibfield  {journal} {\bibinfo
  {journal} {Quantum Information \& Computation}\ }\textbf {\bibinfo {volume}
  {25}},\ \bibinfo {pages} {290} (\bibinfo {year} {2025})}\BibitemShut
  {NoStop}%
\bibitem [{\citenamefont {Ghiu}\ \emph {et~al.}(2020)\citenamefont {Ghiu},
  \citenamefont {Grimaudo}, \citenamefont {Mihaescu}, \citenamefont {Isar},\
  and\ \citenamefont {Messina}}]{ghiu2020quantum}%
  \BibitemOpen
  \bibfield  {author} {\bibinfo {author} {\bibfnamefont {I.}~\bibnamefont
  {Ghiu}}, \bibinfo {author} {\bibfnamefont {R.}~\bibnamefont {Grimaudo}},
  \bibinfo {author} {\bibfnamefont {T.}~\bibnamefont {Mihaescu}}, \bibinfo
  {author} {\bibfnamefont {A.}~\bibnamefont {Isar}},\ and\ \bibinfo {author}
  {\bibfnamefont {A.}~\bibnamefont {Messina}},\ }\bibfield  {title} {\bibinfo
  {title} {Quantum correlation dynamics in controlled two-coupled-qubit
  systems},\ }\href@noop {} {\bibfield  {journal} {\bibinfo  {journal}
  {Entropy}\ }\textbf {\bibinfo {volume} {22}},\ \bibinfo {pages} {785}
  (\bibinfo {year} {2020})}\BibitemShut {NoStop}%
\bibitem [{\citenamefont {Xu}\ and\ \citenamefont
  {Xu}(2012)}]{xu2012protecting}%
  \BibitemOpen
  \bibfield  {author} {\bibinfo {author} {\bibfnamefont {H.-S.}\ \bibnamefont
  {Xu}}\ and\ \bibinfo {author} {\bibfnamefont {J.-b.}\ \bibnamefont {Xu}},\
  }\bibfield  {title} {\bibinfo {title} {Protecting quantum correlations of two
  qubits in independent non-markovian environments<? a3b2 show [pmg: line-break
  justify=" yes"/]?> by bang-bang pulses},\ }\href@noop {} {\bibfield
  {journal} {\bibinfo  {journal} {Journal of the Optical Society of America B}\
  }\textbf {\bibinfo {volume} {29}},\ \bibinfo {pages} {2074} (\bibinfo {year}
  {2012})}\BibitemShut {NoStop}%
\bibitem [{\citenamefont {Ferreira}\ \emph {et~al.}(2021)\citenamefont
  {Ferreira}, \citenamefont {Banker}, \citenamefont {Sipahigil}, \citenamefont
  {Matheny}, \citenamefont {Keller}, \citenamefont {Kim}, \citenamefont
  {Mirhosseini},\ and\ \citenamefont {Painter}}]{ferreira2021collapse}%
  \BibitemOpen
  \bibfield  {author} {\bibinfo {author} {\bibfnamefont {V.~S.}\ \bibnamefont
  {Ferreira}}, \bibinfo {author} {\bibfnamefont {J.}~\bibnamefont {Banker}},
  \bibinfo {author} {\bibfnamefont {A.}~\bibnamefont {Sipahigil}}, \bibinfo
  {author} {\bibfnamefont {M.~H.}\ \bibnamefont {Matheny}}, \bibinfo {author}
  {\bibfnamefont {A.~J.}\ \bibnamefont {Keller}}, \bibinfo {author}
  {\bibfnamefont {E.}~\bibnamefont {Kim}}, \bibinfo {author} {\bibfnamefont
  {M.}~\bibnamefont {Mirhosseini}},\ and\ \bibinfo {author} {\bibfnamefont
  {O.}~\bibnamefont {Painter}},\ }\bibfield  {title} {\bibinfo {title}
  {Collapse and revival of an artificial atom coupled to a structured photonic
  reservoir},\ }\href@noop {} {\bibfield  {journal} {\bibinfo  {journal}
  {Physical Review X}\ }\textbf {\bibinfo {volume} {11}},\ \bibinfo {pages}
  {041043} (\bibinfo {year} {2021})}\BibitemShut {NoStop}%
\bibitem [{\citenamefont {Gyongyosi}(2020)}]{gyongyosi2020decoherence}%
  \BibitemOpen
  \bibfield  {author} {\bibinfo {author} {\bibfnamefont {L.}~\bibnamefont
  {Gyongyosi}},\ }\bibfield  {title} {\bibinfo {title} {Decoherence dynamics
  estimation for superconducting gate-model quantum computers.},\ }\href@noop
  {} {\bibfield  {journal} {\bibinfo  {journal} {Quantum Information
  Processing}\ }\textbf {\bibinfo {volume} {19}} (\bibinfo {year}
  {2020})}\BibitemShut {NoStop}%
\bibitem [{\citenamefont {Gyongyosi}\ and\ \citenamefont
  {Imre}(2021)}]{gyongyosi2021scalable}%
  \BibitemOpen
  \bibfield  {author} {\bibinfo {author} {\bibfnamefont {L.}~\bibnamefont
  {Gyongyosi}}\ and\ \bibinfo {author} {\bibfnamefont {S.}~\bibnamefont
  {Imre}},\ }\bibfield  {title} {\bibinfo {title} {Scalable distributed
  gate-model quantum computers},\ }\href@noop {} {\bibfield  {journal}
  {\bibinfo  {journal} {Scientific reports}\ }\textbf {\bibinfo {volume}
  {11}},\ \bibinfo {pages} {5172} (\bibinfo {year} {2021})}\BibitemShut
  {NoStop}%
\bibitem [{\citenamefont {Gyongyosi}(2022)}]{gyongyosi2022adaptive}%
  \BibitemOpen
  \bibfield  {author} {\bibinfo {author} {\bibfnamefont {L.}~\bibnamefont
  {Gyongyosi}},\ }\bibfield  {title} {\bibinfo {title} {Adaptive problem
  solving dynamics in gate-model quantum computers},\ }\href@noop {} {\bibfield
   {journal} {\bibinfo  {journal} {Entropy}\ }\textbf {\bibinfo {volume}
  {24}},\ \bibinfo {pages} {1196} (\bibinfo {year} {2022})}\BibitemShut
  {NoStop}%
\bibitem [{\citenamefont {Gyongyosi}\ and\ \citenamefont
  {Imre}(2022)}]{gyongyosi2022advances}%
  \BibitemOpen
  \bibfield  {author} {\bibinfo {author} {\bibfnamefont {L.}~\bibnamefont
  {Gyongyosi}}\ and\ \bibinfo {author} {\bibfnamefont {S.}~\bibnamefont
  {Imre}},\ }\bibfield  {title} {\bibinfo {title} {Advances in the quantum
  internet},\ }\href@noop {} {\bibfield  {journal} {\bibinfo  {journal}
  {Communications of the ACM}\ }\textbf {\bibinfo {volume} {65}},\ \bibinfo
  {pages} {52} (\bibinfo {year} {2022})}\BibitemShut {NoStop}%
  @article{berta2010uncertainty,
  title     = {The uncertainty principle in the presence of quantum memory},
  author    = {Berta, Mario and Christandl, Matthias and Colbeck, Roger and Renes, Joseph M. and Renner, Renato},
  journal   = {Nature Physics},
  volume    = {6},
  number    = {9},
  pages     = {659--662},
  year      = {2010},
  publisher = {Nature Publishing Group},
  doi       = {10.1038/nphys1734}
  }
  \bibitem{berta2010uncertainty}
  M. Berta, M. Christandl, R. Colbeck, J. M. Renes, and R. Renner,
  ``The uncertainty principle in the presence of quantum memory,''
  \textit{Nature Physics} \textbf{6}, 659--662 (2010),
  doi: \href{https://doi.org/10.1038/nphys1734}{10.1038/nphys1734}.
  
  
\end{thebibliography}%
\end{document}